\documentclass[a4paper,11pt]{article}
\pdfoutput=1 
\RequirePackage{snapshot}  

\usepackage{jheppub} 

\usepackage{bm,amsmath,amssymb,slashed,graphicx,%
            enumerate,alltt,xspace,multirow,xcolor,mathrsfs}
\usepackage{fancyvrb}
\usepackage{booktabs,tabularx}
\usepackage{graphicx}
\usepackage{subcaption}
\usepackage{xspace}
\usepackage{cancel}
\usepackage[utf8]{inputenc} 
\usepackage[compat=1.0.0]{tikz-feynman}
\usepackage{scalerel}
\usepackage{layouts}
\usepackage{adjustbox}
\usepackage{pdflscape}
\usepackage[normalem]{ulem} 

\usepackage{printlen}
\usepackage{wasysym}

\newcommand{\re}{{\rm{Re}}}
\newcommand{\Tr}{{\rm Tr}}

\def\q{{\boldsymbol q}}
\def\0{{\boldsymbol 0}}
\def\1{{\boldsymbol 1}}
\def\p{{\boldsymbol p}}
\def\pbar{{\boldsymbol{\bar p}}}
\def\bkappa{{\boldsymbol{\kappa}}}
\def\bkappabar{{\boldsymbol{\bar \kappa}}}
\def\l{{\boldsymbol l}}
\def\k{{\boldsymbol k}}

\def\r{{\boldsymbol r}}
\def\r{{\boldsymbol r}}

\def\v{{\boldsymbol v}}
\def\w{{\boldsymbol w}}

\def\0{{\boldsymbol 0}}

\newcommand{\ii}{\mathrm{i}\,}

\renewcommand\d{\delta}

\newcommand\m{\mu}

\newcommand\s{\sigma}

\def\cV{\mathcal{V}}

\def\cG{{\cal G}}

\def\cA{{\cal A}}

\def\cM{{\cal M}}

\def\cW{{\cal W}}

\def\Tr{\text{Tr}}

\usepackage[capitalise]{cleveref}
\makeatletter
\g@addto@macro\bfseries{\boldmath}
\makeatother

\definecolor{labelkey}{rgb}{0,0.5,0.0}
\definecolor{royalpurple}{rgb}{0.47, 0.32, 0.66}

\usepackage{listings}
\definecolor{darkgreen}{rgb}{0,0.4,0}
\definecolor{grey}{rgb}{0.5,0.5,0.5}
\definecolor{rust}{rgb}{0.9,0.4,0.0}

\allowdisplaybreaks 

\newcommand{\Fmed}{F_{\rm{med}}}
\newcommand{\TildeDeltamed}{\widetilde{\Delta}_{\rm med}}
\newcommand{\Deltamed}{\Delta_{\rm{med}}}
\newcommand{\qbar}{\bar q}

\title{A generalized picture of colour decoherence 
in dense QCD media}

\preprint{CERN-TH-2024-185}

\newcommand{\CERNaff}{CERN, Theoretical Physics Department, CH-1211 Geneva 23, Switzerland}
\newcommand{\IGFAEaff}{Instituto Galego de F\'{i}sica de Altas Enerx{\'{i}}as,  Universidade de Santiago de Compostela, Santiago de Compostela 15782,  Galicia, Spain}
\newcommand{\LIPaff}{Laborat{\'{o}}rio de Instrumenta\c{c}ão e F{\'{i}}sica Experimental de Part{\'{i}}culas (LIP),
Av.~Prof.~Gama Pinto, 2, P-1649-003 Lisboa, Portugal}
\newcommand{\ISTaff}{Departmento de F\'{i}sica, Instituto Superior T\'{e}cnico (IST),
Universidade de Lisboa, Av.~Rovisco Pais 1, P-1049-001 Lisboa, Portugal}
\newcommand{\HIGGSaff}{Higgs Centre for Theoretical Physics, School of Physics and Astronomy, The University of Edinburgh, Edinburgh EH9 3FD, Scotland, United Kingdom}
\newcommand{\UGRaff}{Departamento de F\'{i}sica Te\'{o}rica y del Cosmos, Universidad de Granada,
Campus de Fuentenueva, E-18071 Granada, Spain}

\author[a,b]{Samuel Abreu,}%
\author[c]{Xo{\'{a}}n Mayo L\'{o}pez,}%
\author[d,e]{Guilherme Milhano,}%
\author[f]{Alba Soto-Ontoso}%

\affiliation[a]{\CERNaff}
\affiliation[b]{\HIGGSaff}
\affiliation[c]{\IGFAEaff}
\affiliation[d]{\LIPaff}
\affiliation[e]{\ISTaff}
\affiliation[f]{\UGRaff}

\date{Received: date / Accepted: \today}

\abstract{
  We revisit the calculation of the soft gluon emission probability off a colour-singlet $q\bar q$ system that evolves in a quark-gluon plasma. The $q\bar q$ antenna is created in the presence of a medium and then emits a soft gluon outside.
  The gluon emission probability is modified with respect to the vacuum baseline due to interactions with the medium during the formation of the antenna and its propagation. Previous studies disregarded the former effect and found that the medium modification to the interference pattern of the antenna was controlled by the so-called critical angle $\theta_c$, that exclusively depends on medium properties. We find that accounting for medium interactions during the antenna formation enhances the total rate of emissions off the $q\bar q$ antenna. Interestingly, it also promotes the notion of a critical angle to a dynamic quantity, denoted $\tilde\theta_c$, that depends on both the medium and the antenna properties and is thus different for every splitting. 
  As a consequence, depending on the region of parameter space, colour decoherence can either be delayed or accelerated with respect to previous estimates. 
}
\keywords{QCD, QGP, parton branching, colour coherence
  \\[4em]
  \textit{For the purpose of Open Access, the authors have applied a CC BY
  public copyright licence to any Author Accepted Manuscript (AAM)
  version arising from this submission.}
}

\begin{document}

\maketitle

\section{Introduction}
\label{sec:introduction}
Jet physics plays a key role in the characterization of the hot and dense coloured medium created in heavy-ion collisions known as the quark-gluon plasma (QGP)~\cite{Majumder:2010qh,Mehtar-Tani:2013pia,Blaizot:2015lma,Connors:2017ptx,Cunqueiro:2021wls,Apolinario:2022vzg,Apolinario:2024equ}. One of the most robust experimental signatures of QGP formation is the modification of the jet transverse momentum ($p_t$) spectrum with respect to proton-proton collisions ($pp$). Namely, the cross-section for producing a highly energetic jet with a given cone size $R$ in a heavy-ion collision is suppressed~\cite{ATLAS:2012tjt,STAR:2020xiv,CMS:2021vui,ALICE:2023waz}, with, for instance, a $30\%$ suppression being observed for central $\rm{PbPb}$ collisions at $p_t=400$ GeV and $R=0.4$~\cite{CMS:2021vui}. 

The depletion of the jet spectrum is closely related to the angular structure of the single emission matrix-element in QCD. In the absence of a medium, the probability of radiating at small angles is logarithmically enhanced because the matrix element contains a collinear singularity. In the medium, colour and momentum exchanges with the QGP, both during the formation time of the emission and its propagation, lead to a finite emission probability in the collinear limit~\cite{Baier:1996kr,Zakharov:1996fv,Gyulassy:1999zd,Wiedemann:2000za}. Consequently, medium-induced emissions are radiated preferably at wide angles and broadened away from the emitter's direction by subsequent interactions with the medium~\cite{Casalderrey-Solana:2010bet}. Some of this radiation might end up outside of the jet-cone of radius $R$, leading to some $p_t$ being lost when reconstructing the jet, and thus inducing a shift in the spectrum compared to the $pp$ result. 

A natural question is whether this simple explanation for the suppression of high-$p_t$ jets still holds when accounting for subsequent emissions. For instance, it is well known that the angular structure of the QCD cascade in vacuum is dictated by quantum mechanical interference effects~\cite{Dokshitzer:1982fh,Webber:1983if,Marchesini:1987cf,Dokshitzer:1987nm}. This can be understood at the level of two emissions by calculating the soft-gluon emission probability off a $q\bar q$ antenna in a colour singlet configuration. Considering the emission probability from the quark leg one finds (see e.g.~\cite{Dokshitzer:1991wu,Ellis:1996mzs})
\begin{equation}
\label{eq:angular-ord}
d N^{{\rm vac}}_q=
\frac{\alpha_s C_F}{\pi}\frac{dE_g}{E_g}
\frac{\sin\theta_g\,d\theta_g}{1-\cos\theta_g}
\Theta(\theta_{q\bar q}-\theta_g)\,,
\end{equation}
where $\theta_{q\bar q}$ denotes the angle between the two prongs of the antenna, and $\theta_g$ the angle between the gluon and the quark.
This result indicates that soft radiation at angles larger than the antenna's opening angle is forbidden.\footnote{For a non-singlet configuration, Eq.~\eqref{eq:angular-ord} receives another contribution to the spectrum proportional to $C_A \Theta(\theta_g-\theta_{q\bar q})$ that can be interpreted as the large-angle gluon emissions triggered by the parent gluon.} Diagrammatically, this angular structure arises from the interplay between the contributions where the gluon is emitted and reabsorbed by the (anti-)quark in the amplitude and conjugate amplitude, and those
where the gluon is emitted by the quark in the amplitude and absorbed by the anti-quark in the conjugate amplitude.
Thus, interference effects are responsible for the angular-ordering property of QCD radiation in vacuum, often called colour coherence.

The soft-gluon emission probability off a $q\bar q$ antenna including in-medium effects was computed almost 15 years ago in a series of works~\cite{Mehtar-Tani:2010ebp,Mehtar-Tani:2011hma,Casalderrey-Solana:2011ule,Mehtar-Tani:2011vlz,Mehtar-Tani:2011lic,Armesto:2011ir,Mehtar-Tani:2012mfa,Casalderrey-Solana:2012evi,Calvo:2014cba}. The theoretical setup considered a $q\bar q$ antenna that propagates through a brick of QGP with length $L$, emitting a soft gluon either inside or outside the medium. For simplicity, and because it already captures the main dynamical effects, we will only discuss the case in which the gluon is emitted outside. Physically, this corresponds to gluon emissions with long formation times\footnote{The concept of formation time is subtle, and we will return to it in the rest of the paper. To be more explicit here, let us  denote the opening angle of a dipole as $\theta$. The prongs have transverse momenta/energies $p_{t1}$ and $p_{t2}$. Then, the vacuum formation time of an emission is given by $t_f=2/(k_t\theta)$, with $k_t=\min(p_{t1},p_{t2})\theta$.} 
compared to the length of the medium, and therefore very soft. Regarding the formation of the antenna, the authors considered an instantaneous formation in the medium, which effectively sets to zero the light-cone time difference $\Delta x^+_A$ between the splitting in the amplitude and conjugate amplitude. 
Within this setup, the squared matrix element for gluon emission off the antenna was found to be
\begin{equation}
\label{eq:anti-angular-ord}
    \mathcal{M}^2_{q\bar q g} 
    \sim\left(\frac{1}{\bkappa^2}+\frac{1}{\bkappabar^2}
    -2
    \frac{\bkappa \cdot \bkappabar}{\bkappa^2 \bkappabar^2}\right) +2
    \frac{\bkappa \cdot \bkappabar}{\bkappa^2 \bkappabar^2} \Deltamed ,
\end{equation}
where the first term in brackets is the vacuum result  leading to the angular-ordered radiation described by Eq.~\eqref{eq:angular-ord}, and the term proportional to $\Deltamed$ leads to anti-angular-ordered radiation in the complementary angular region, that is in the region corresponding to $\Theta(\theta_g-\theta_{q\bar q})$.
The strength of this radiation is controlled by the value of $\Deltamed$, the so-called `decoherence factor', which accounts for medium effects and is such
that $\Deltamed\to0$ for vacuum and $\Deltamed\to1$
for completely opaque media. 
Eq.~\eqref{eq:anti-angular-ord} unveils a region of phase-space, controlled by medium properties, for which vacuum-like emissions at angles larger than $\theta_{q\bar q}$ are allowed. Thus, interference effects enable anti-angular ordered medium-induced emissions.

The physical mechanism underlying this striking observation is called colour decoherence, and it relies on the fact that, to first approximation, the parton-medium interaction is a colour rotation. Therefore, the antenna, which starts its propagation in a colour singlet configuration, will undergo multiple interactions with the QGP
which amount to colour rotations,
and eventually the prongs become de-correlated in colour space. Effectively, after interacting sufficiently with the medium, the quark and anti-quark will behave as independent colour charges and the radiation phase-space is thus unconstrained. This is known as the total decoherence regime~\cite{Mehtar-Tani:2011hma}, and corresponds to setting $\Deltamed=1$ in Eq.~\eqref{eq:anti-angular-ord}, canceling the
interference contribution.

The previous discussion disregards the medium-modification to the antenna formation itself. Indeed, the cross-section to create an antenna is modified with respect to vacuum when
allowing for a (light-cone) time gap $\Delta x_A^+$ 
between its formation in amplitude and conjugate amplitude~\cite{Dominguez:2019ges,Isaksen:2023nlr}.
Owing to this interference effect, the in-medium antenna cross-section can be written as
\begin{equation}
\label{eq:fmed-basic}
\frac{d\sigma}{d z_q d\theta_{q\bar q}} = \frac{d\sigma^{\rm vac}}{d z_q d\theta_{q\bar q}}[1+F_{\rm {med}}(E_\gamma, z_q,\theta_{q\bar q})],
\end{equation}
where $E_\gamma$ is the energy of the parent photon, $z_q$ is the energy-sharing fraction between the antenna prongs, and $\theta_{q\bar q}$ the opening angle.
The medium-modification factor $\Fmed$ vanishes
in the case of no medium. As one would expect, Eq.~\eqref{eq:fmed-basic}  reduces
to the vacuum expression when $\Delta x_A^+=0$ (even though
this limit is more subtle to take). This effect
has been studied in 
Refs.~\cite{Dominguez:2019ges} and \cite{Isaksen:2023nlr}. 
The former used a semi-classical treatment, while the latter included finite $N_c$ and finite $z_q$ corrections. These studies have shown that $\Fmed$ deviates from unity for a colour-singlet antenna $\gamma\to q\bar q$ for unbalanced splittings ($z_q\ll 0.5$) at wide angles. Interestingly, this indicates that hard-collinear splittings remain vacuum-like even in the presence of a medium, as was also pointed out in Ref.~\cite{Caucal:2018dla}. 

In this work, we calculate the soft-gluon emission probability off a QCD antenna in a colour singlet configuration, including both medium interference effects discussed above. That is, we account for interactions during the antenna propagation through a medium of length $L$ while considering a non-zero $\Delta x^+_A$. Consistently with the fact that the emitted gluon is much less energetic than the antenna legs, we focus on the case in which the emission takes place outside the medium. The main result of this paper is
the squared matrix element for gluon emission off a colour
singlet antenna, accounting for non-zero $\Delta x^+_A$ effects,
which is given by 
\begin{equation}
\label{eq:final-result-intro}
    \cM_{q\bar q g}^2\sim(1+\Fmed)\left[\left(\frac{1}{\bkappa^2}+\frac{1}{\bkappabar^2}
    -2
    \frac{\bkappa \cdot \bkappabar}{\bkappa^2 \bkappabar^2}\right) +2
    \frac{\bkappa \cdot \bkappabar}{\bkappa^2 \bkappabar^2} \TildeDeltamed \right].
\end{equation}
The generalized decoherence factor $\TildeDeltamed(E_\gamma, z_q,\theta_{q\bar q})$ is controlled by both medium and antenna properties. $\Fmed$ is the same medium modification factor as in Eq.~\eqref{eq:fmed-basic}.
In the vacuum limit, both $\Fmed$ and $\TildeDeltamed$ vanish and Eq.~\eqref{eq:final-result-intro} reproduces the vacuum result. When setting $\Delta x_A^+=0$, Eq.~\eqref{eq:final-result-intro} reduces to Eq.~\eqref{eq:anti-angular-ord} as discussed later in the paper. 
Unlike $\Deltamed$, $\TildeDeltamed(E_\gamma, z_q,\theta_{q\bar q})$ has an intricate functional dependence on $E_\gamma$, $z_q$ and $\theta_{q\bar q}$, and we find that it can be either smaller or larger than $\Deltamed$, thus reducing or enhancing the anti-angular ordered contribution. 
Finally, compared to Eq.~\eqref{eq:anti-angular-ord} the total rate of emissions is enhanced by the $\Fmed$ factor.

The rest of this paper is organized as follows. 
In Section~\ref{sec:theory} we introduce the building blocks of the calculation together with the approximations we adopted. The squared matrix element for $\gamma^*\to q\bar q g$ is computed in Section~\ref{sec:me},
where we also define the generalized decoherence factor 
$\TildeDeltamed$. In Section~\ref{sec:numerics}, we present a numerical study of our results, comparing them to the $\Deltamed$ baseline.
We conclude with some future prospects and discussion of phenomenological applications of this study in Section~\ref{sec:conclusions}. 
In Appendix~\ref{app:bonus} we include plots to illustrate the behaviour of $\Fmed$, as well as a Lund plane representation of $\TildeDeltamed$.

\section{Amplitudes for $\gamma^*\to q\bar q g$ in a dense medium}
\label{sec:theory}

We consider the creation and propagation of a highly boosted
ultra-relativistic $q\bar q$ antenna in the presence of a finite, dense medium.
Each prong of the antenna, with momentum $p$ for the quark and 
$\bar p$ for the anti-quark,
is highly boosted in the $z$ direction, that is $p_0\sim p_z$
and similarly for $\bar p$. It is thus convenient to introduce
light-cone coordinates\footnote{\label{foot:LCvars}The light-cone coordinates are defined as 
$v^{\pm}=\frac{v^0\pm v^z}{\sqrt{2}}$, 
$\v=(v^x,v^y)$,  with the dot product of two 
4-vectors becoming $v\cdot w = v^+ w^- + v^- w^+ - \v\cdot\w$. 
} such that $p=(p^+,p^-,\p)$, where $p^+\gg |\p|,p^-$.
The antenna emits a gluon of momentum $k$, which we take to be very soft.
As it has been argued in the introduction and will be explicitly seen  below, 
this implies that its formation time is very large compared to the length of the medium, and therefore it is emitted outside of it.

Let us start by collecting the variables that we will be using throughout this
paper and comment on their relative sizes. The quark and anti-quark light-cone energies
are denoted $E_q$ and $E_{\bar q}$ (i.e., $E_q\equiv p_+$ and 
$E_{\bar q}\equiv \bar p_+$), and
since the gluon is very soft its energy $E_g$ satisfies $E_g\ll E_q,\,E_{\bar{q}}$.
The total energy is $E_\gamma=E_q+E_{\bar q}+E_g$. In intermediate steps, before taking the aforementioned limits, we keep explicit the sum of the
quark (anti-quark) and gluon energies, which we write as
\begin{equation}
    E_l=E_q+E_g\,,\qquad\qquad E_{\bar l}=E_{\bar q}+E_g\,.
\end{equation}
Finally, we introduce energy fractions which are particularly convenient for
expressing our final result. We have the gluon relative energy,
\begin{equation}
    z_g=\frac{E_g}{E_l}\sim\frac{E_g}{E_q}\,,\qquad
    \bar{z}_g=\frac{E_g}{E_{\bar l}}\sim\frac{E_g}{E_{\bar{q}}}\,,
\end{equation}
and the quark and anti-quark relative energies,
\begin{equation}
    z_q=\frac{E_q+E_g}{E_q+E_{\bar q}+E_g}\sim\frac{E_q}{E_\gamma}\,,\qquad
    \bar{z}_q=\frac{E_{\bar q}+E_g}{E_q+E_{\bar q}+E_g}\sim\frac{E_{\bar q}}{E_\gamma}\,.
\end{equation}
These variables are not all independent since
$\bar{z}_q=1-z_q$ and $\bar{z}_q\bar{z}_g=z_qz_g$, but 
introducing them is convenient to write more 
compact expressions.

The medium has a length $L\equiv L^+$, and it is modelled by a classical colour field, generated by static 
quasi-particle sources which are unaffected by interactions with the 
traversing particles. Therefore, the high-energy partons interact with a stochastic background field which can be written as 
\begin{equation} \label{eq:amu-field}
    g A^{a, \m}_{\text{ext}} (q)= g^{\m+} \int_{x^+} e^{\ii \, q^-x^+} \, \cA^a(x^+,\q) \,(2\pi)\d(q^+) \, ,
\end{equation}
where $g^{\mu\nu}$ is the metric tensor. $\cA^a(x^+,\q)$ contains the information about both the 
colour density of the medium scattering centres and the potential
that controls the interaction between the traversing partons and 
the medium quasi-particles. We note that since the partons are 
highly boosted in the $+$ direction, the static medium sources 
are effectively boosted in the opposite direction and the 
field in Eq.~\eqref{eq:amu-field} only has a non-zero component in
the $-$ direction.

When computing an observable, the stochastic background field 
must be averaged over all possible configurations. 
We assume the field to have Gaussian statistics, i.e., with only 
pairwise averages being non-zero~\cite{McLerran:1993ni,Gyulassy:1993hr}.
Furthermore, we assume that interactions between the partons 
and the medium are local, and that there are no correlations 
between different sources. The average over the field 
configurations is then given by 
\begin{align}\label{eq:medium-correlator}
    \langle \cA^a(x^+,\p) \cA^{*b}(\bar{x}^+,\bar{\p})\rangle = \frac{\d^{ab}}{2C_{\bar R}} \, \rho(x^+) \, \d(x^+-\bar{x}^+) (2\pi)^2 \, \d^{(2)}(\p-\bar{\p}) \, \cV(\p) \, ,
\end{align}
where $\rho(x^+)$ is the number density of in-medium sources, 
$C_{\bar R}=C_A$ for quarks and $C_{\bar R}=C_F$ for gluons 
interacting with the medium, and $\cV(\p)$ is a screened Coulomb-like 
potential that controls the interaction between the traversing
partons and the medium. Corrections to this simple description of the medium incorporating inhomogeneities and flow effects
are the subject of an active field of research~\cite{Sadofyev:2021ohn,Hauksson:2021okc,Antiporda:2021hpk,Barata:2022krd,Andres:2022ndd,Fu:2022idl,Barata:2023qds,Barata:2023zqg,Kuzmin:2023hko,Kuzmin:2024smy}.

\begin{figure}
    \begin{subfigure}[]{0.49\textwidth}
    \centering
    \includegraphics[width=\textwidth]{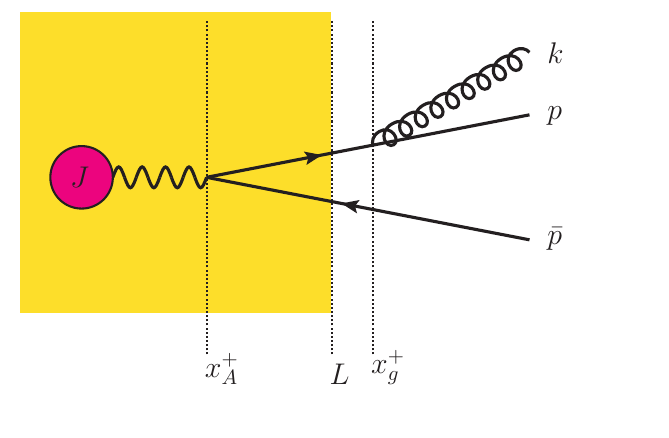}
    \caption{{\bf in}: Antenna formation inside the medium}
    \label{fig:mqIn}
    \end{subfigure}
    \begin{subfigure}[]{0.49\textwidth}
    \centering
    \includegraphics[width=\textwidth]{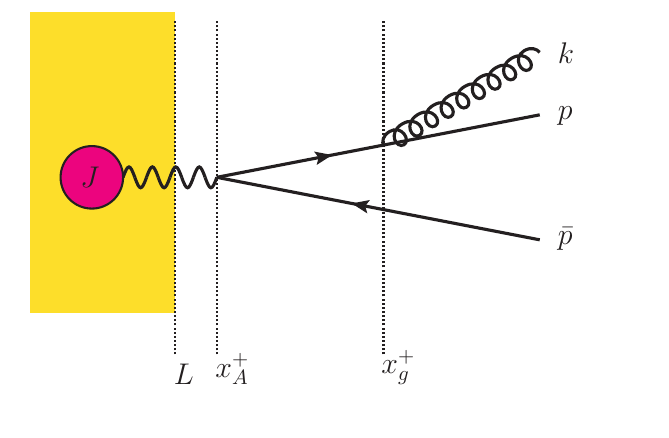}
    \caption{{\bf out}: Antenna formation outside the medium}
    \label{fig:mqOut}
    \end{subfigure}
    \caption{Finite medium, with gluon emission outside the medium}
    \label{fig:mqFiniteMedium}
\end{figure}

\begin{figure}\centering
    \includegraphics[width=.5\textwidth]{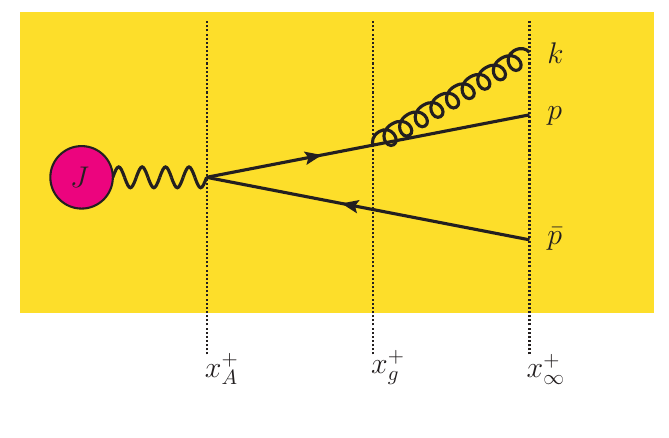}
    \caption{Antenna formation and gluon emission inside an infinite medium}
    \label{fig:infiniteMedium}
\end{figure}

Given our assumption that the gluon is very soft and 
emitted outside the medium,
there are two different amplitudes we should consider 
(see Fig.~\ref{fig:mqFiniteMedium}):
the one where the antenna is formed inside the medium (denoted {\bf in}),
and the one where the antenna is formed outside
(denoted {\bf out}). That is, when the gluon is emitted by
the quark we write
\begin{equation}
\label{eq:mq-in-out-split}
    M_q = M^{{\bf in}}_q + M^{{\bf out}}_q \,,
\end{equation}
and similarly for $M_{\bar q}$, when the gluon is emitted off
the anti-quark line.
For simplicity, however, we will first write
the amplitude for the process where the antenna formation
and the gluon emission both happen inside an infinite medium
(see Fig.~\ref{fig:infiniteMedium}), 
and then take the necessary limits of this expression.

The Feynman rules describing the propagation of
particles inside a medium as described above are well known and have been widely used. They can be found
in e.g.~Refs.~\cite{Mehtar-Tani:2017ypq,Barata:2024bqp}. 
With these Feynman rules, 
the amplitude corresponding to Fig.~\ref{fig:infiniteMedium} is:
\begin{align} 
\begin{split}\label{eq:mqMedium}
    \ii M^{\rm med}_q =& - \frac{g\, e}{4 E_l E_\gamma} \, \int_{0}^{x^+_\infty} dx^+_g \int_{0}^{x^+_g} dx^+_A \int_{\l_1,\k_1,\p_1,\pbar_1} 
    e^{\ii\frac{\k^2}{2E_g}x^+_\infty} \cG^{bb_1}\left(x^+_\infty,\k; x^+_g,\k_1\right)\\
    &e^{\ii\frac{\p^2}{2E_q}x^+_\infty} \cG_{ii_2}\left(x^+_\infty,\p;x^+_g,\p_1\right)
    t^{b_1}_{i_2i_1}\gamma^{\s_q, \s}_{g,\lambda_g}\left[p_1;k_1\right]
    \cG_{i_1i_A}\left(x^+_g,\p_1+\k_1;x^+_A,\l_1\right)\\
    &\delta_{i_Aj_A}\gamma^{\s,\s_{\bar{q}}}_{A,\lambda_A}\left[l_1;\bar p_1\right]
    \epsilon^{*\lambda_A}\left(l_1+\bar p_1\right)\cdot J\left(l_1+\bar p_1\right)
    e^{-\ii\frac{(\l_1+\pbar_1)^2}{2E_\gamma} x^+_a}\\
    &e^{\ii\frac{\bar{\p}^2}{2 E_{\bar{q}}} x^+_\infty}\bar{\cG}_{j_Aj}\left(x^+_\infty,\bar{\p};x^+_a,\bar\p_1\right)\,.
\end{split}
\end{align}
where $g^2= 4\pi \alpha_s$ and $e$ is the electric charge. 
Let us make some comments on this expression. 
The propagation of each particle inside the medium is described
by an in-medium propagator, $\cG$, which allows for colour rotations and transverse motion
as particles advance through the matter. 
In our approximation, the spinor index or helicity 
of the particles is not modified by medium interactions.
The external spinors and polarisations are absorbed into the definition
of the medium propagators, 
and the vertices include the spinors and helicities
at the interaction point. More explicitly, the vertices are
\begin{align}\begin{split}\label{eq:verticesAmp}
    &\gamma^{\s,\s_{\bar{q}}}_{A,\lambda_A}\left[{l}_1;\bar{p}_1\right] = 
    \bar{u}^\s \left({l}_1\right) \slashed{\epsilon}^{\lambda_A} \left({l}_1+\bar{p}_1\right) 
    v^{\s_{\bar{q}}}\left(\bar{p}_1\right) \, ,
    \\ & \gamma^{\s_q, \s}_{g,\lambda_g}\left[{p}_1;{k}_1\right] = \bar{u}^{\s_q}\left({p}_1\right) \slashed{\epsilon}^{*\lambda_g}\left({k}_1\right) u^\s\left({p}_1+{k}_1\right) \, ,
\end{split}\end{align}
where $\s_q$ and $\s_{\bar{q}}$ are the spinor indices of the external quark and anti-quark, and $\lambda_g$
the polarisation of the external gluon.
The current $J^\mu$ describes the creation of the
virtual photon.
Note that these expressions only depend on the energy and transverse components 
of each of the momenta, since the minus component has been integrated out introducing the dependence on the light-cone time, but for simplicity of the notation we write them as depending on the
full four-momenta.
Finally, in Eq.~\eqref{eq:mqMedium} there is a phase associated with each external propagator, as well as the one corresponding to the vacuum propagator of the photon.

Given that the prongs of the antenna are very energetic, 
we assume that they propagate along straight
lines and we write the quark propagators as (tilted) eikonal
propagators. Thus, the description of highly unbalanced splittings, with $z_q$ very close to 0 or 1, is beyond the regime of validity of our approximation. Following Refs.~\cite{Altinoluk:2014oxa,Dominguez:2019ges}, if we further set the initial position of the eikonal 
propagators in the transverse plane to zero, then we have that
\begin{equation}
    \label{eq:eikProp}
    \cG^{\text{eik}}_{ij}\left(x_2^+,\p_2;x_1^+,\p_1\right)=
    \,(2\pi)^2 \, \d^{(2)}(\p_2-\p_1) \, e^{-\ii\frac{\p_2^2}{2E}(x^+_2-x^+_1)}
    \cW_{ij}\left(x^+_2,x^+_1; \frac{\p_2}{E}(\tau-x^+_1)\right)\,,
\end{equation}
and similarly for the anti-quark. Replacing the propagators in Eq.~\eqref{eq:mqMedium} by eikonal propagators already leads to a considerable simplification
that allows us to perform some of the integrations over transverse components in the amplitude. 
To further simplify it, we focus on the limits 
corresponding to Fig.~\ref{fig:mqFiniteMedium}. We start with the {\bf in}
case of Fig.~\ref{fig:mqIn}. Taking the amplitude in Eq.~\eqref{eq:mqMedium}
as our starting point, we note that the emitted gluon is outside
the medium, so its propagator becomes a free propagator,
\begin{equation}
    \cG^{bb_1}\left(x^+_\infty,\k; x^+_g,\k_1\right)
    \to
    \delta_{bb_1}(2\pi)^2 \, \d^{(2)}(\k_1-\k) \, e^{-\ii\frac{\k^2}{2E_g}(x^+_\infty-x^+_g)}\,.
\end{equation}
Combined with the replacement of the quark and anti-quark propagators 
with eikonal propagators, this allows us to perform all the transverse 
momentum integrals.
We also note that after these replacements
the dependence on $x_\infty^+$ in the phases cancels out.
Finally, we note that the Wilson lines in the quark and anti-quark 
eikonal propagators either simplify or become `shorter'.
Explicitly,
\begin{align}\begin{split}
    &\cW_{ii_2}\left(x^+_\infty,x^+_g;\frac{\p}{E_q}(\tau-x^+_g)\right)\to\delta_{ii_2}\,,\\
    &\cW_{i_1i_A}\left(x^+_g,x^+_A;\frac{\p+\k}{E_q+E_g}(\tau-x^+_g)\right)\to
    \cW_{i_1i_A}\left(L,x^+_A;\r_{q}^a(\tau)\right)\,,\\
    &\cW^\dagger_{j_Aj}\left(x^+_g,x^+_A;\frac{\bar\p}{E_{\bar q}}(\tau-x^+_g)\right)\to
    \cW^\dagger_{j_Aj}\left(L,x^+_A;\r_{\bar{q}}^A(\tau)\right)\,,
\end{split}\end{align}
where for the quark line we also neglected the gluon momentum compared to the 
much harder quark momentum, and we have defined the trajectories
\begin{equation}
\label{eq:w-line-traj-q}
    \r_{q}^A(\tau) = \frac{\p}{E_q}(\tau-x^+_A)\,,  \qquad\qquad 
    \r_{\bar{q}}^A(\tau)= \frac{\bar{\p}}{E_{\bar{q}}} (\tau-x^+_A)\,.
\end{equation}
With these simplifications and assumptions, the momenta at each vertex 
become the external momenta of the quark, anti-quark and gluon.

Finally, we obtain the amplitude corresponding to the
diagram in Fig.~\ref{fig:mqIn}:
\begin{align} \begin{split}\label{eq:mqIn_temp}
    \ii M^{{\bf in}}_q =& - \frac{g\, e}{4 E_l E_\gamma} 
    \int_{L}^{x^+_\infty} dx^+_g e^{\ii\frac{x^+_g}{t_g}} 
    \int_{0}^{L} dx^+_A e^{\ii \frac{x^+_A}{t_A}} 
    t^{b}_{ii_1}\, \cW_{i_1i_A}\left(L,x^+_A; \r_{q}^A(\tau)\right)
    \\ & \times 
    \cW_{i_Aj}^\dag\left(L,x^+_A;\r_{\bar{q}}^A(\tau)\right) \gamma^{\s_q, \s}_{g,\lambda_g}\left[p;k\right] \, 
    \gamma^{\s,\s_{\bar{q}}}_{A,\lambda_A}[p;\bar{p}] \, \epsilon^{*\lambda_A}\left({p}+{\bar{p}}\right)\cdot J\left({p}+{\bar{p}}\right)
    \,,
\end{split}\end{align}
where we took the limit of a soft gluon and 
we introduced the formation times for the gluon and the antenna,
\begin{equation}\label{eq:formation-times}
    t_{g} = \frac{2 z_g z_q E_\gamma}{(\k-z_g\p)^2} \,,\qquad\qquad
    t_{A} = \frac{2z_q(1-z_q)E_\gamma}{[(1-z_q)\p-z_q\bar{\p}]^2}\,.
\end{equation}
As argued previously, it is clear that for a very soft gluon
(i.e., $z_g,|\k|\to0$) the gluon formation time is very large.

Given that in the diagram of Fig.~\ref{fig:mqIn} the point $x^+_g$ is an arbitrary
point in vacuum, we would expect to be able to trivially integrate it, and this is indeed the case
in Eq.~\eqref{eq:mqIn_temp} as it only appears in a phase. The contribution of
this integral at $x^+_\infty$ vanishes, while the contribution at the other limit of integration gives a phase $e^{\ii L/t_g}$, which we can set to 1 because $t_g\gg L$ for a very soft gluon. 
We then find
\begin{align}\begin{split}\label{eq:mqIn}
    \ii M^{{\bf in}}_q =&  -\frac{\ii\,g\, e\, t_g}{4 E_l E_\gamma} 
    \int_{0}^{L} dx^+_A e^{\ii \frac{x^+_A}{t_A}} 
    t^{b}_{ii_1}\, \cW_{i_1i_A}\left(L,x^+_A; \r_{q}^A(\tau)\right) \cW_{i_Aj}^\dag\left(L,x^+_A;\r_{\bar{q}}^A(\tau)\right) %
    \\ & \gamma^{\s_q, \s}_{g,\lambda_g}\left[p;k\right] \, 
    \gamma^{\s,\s_{\bar{q}}}_{A,\lambda_A}[p;\bar p] \, 
    \epsilon^{*\lambda_A}\left({p}+{\bar{p}}\right)\cdot J\left({p}+{\bar{p}}\right)\,.
\end{split}\end{align}

The {\bf out} case, corresponding to Fig.~\ref{fig:mqOut}, is trivial to obtain following exactly the same steps. We get
\begin{align}\begin{split}\label{eq:mqOut_temp}
    \ii M^{{\bf out}}_q =& - \frac{g\, e}{4 E_l E_\gamma} 
    \int_{L}^{x^+_\infty} dx^+_g e^{\ii\frac{x^+_g}{t_g}} \int_{L}^{x^+_g} dx^+_A e^{\ii \frac{x^+_A}{t_A}}  
    \\ & t^{b}_{ij}\gamma^{\s_q, \s}_{g,\lambda_g}\left[p;k\right] \, 
    \gamma^{\s,\s_{\bar{q}}}_{A,\lambda_A}[p;\bar{p}] \, 
    \epsilon^{*\lambda_A}\left({p}+{\bar{p}}\right)\cdot J\left({p}+{\bar{p}}\right)
    \,.
\end{split}\end{align}
Noticing that both the $x^+_g$ and $x_A^+$  dependences can be integrated trivially we find
\begin{align}\begin{split}\label{eq:mqOut}
    \ii M^{{\bf out}}_q =& \frac{g\, e}{4 E_l E_\gamma} 
    t_A\,t_g e^{\ii \frac{L}{t_A}}
    t^{b}_{ij}\,\gamma^{\s_q, \s}_{g,\lambda_g}\left[p;k\right] \, 
    \gamma^{\s,\s_{\bar{q}}}_{A,\lambda_A}[p;\bar{p}] \,\epsilon^{*\lambda_A}\left({p}+{\bar{p}}\right)\cdot J\left({p}+{\bar{p}}\right)
    \,,
\end{split}\end{align}
where we have again used that $t_g\gg L$.

The amplitudes corresponding to emissions off the anti-quark leg
can be obtained in exactly the same way. We get:
\begin{align}\begin{split}\label{eq:mqbarIn}
    \ii M^{{\bf in}}_{\bar q} =&  \frac{\ii\,g\, e\, \bar t_g}{4 E_{\bar l} E_\gamma} 
    \int_{0}^{L} dx^+_A e^{\ii \frac{x^+_A}{t_A}} 
    \cW_{ii_A}\left(L,x^+_A; \r_{q}^A(\tau)\right) \cW_{i_Aj_1}^\dag\left(L,x^+_A;\r_{\bar{q}}^A(\tau)\right)t^{b}_{j_1j}\,  %
    \\ & \bar{\gamma}^{\s,\s_{\bar q}}_{g,\lambda_g}\left[\bar p;k\right] \, 
    \gamma^{\s_{{q}},\s}_{A,\lambda_A}[p;\bar{p}] \, 
    \epsilon^{*\lambda_A}\left({p}+{\bar{p}}\right)\cdot J\left({p}+{\bar{p}}\right)\,,
\end{split}\end{align}
and
\begin{align}\begin{split}\label{eq:mqbarOut}
    \ii M^{{\bf out}}_{\bar q} =&  -\frac{g\, e}{4 E_{\bar l} E_\gamma} 
    t_A\,\bar{t}_g  e^{\ii \frac{L}{t_A}}
    t^{b}_{ij}\,
    \bar{\gamma}^{\s,\s_{\bar q}}_{g,\lambda_g}\left[\bar p;k\right] \, 
    \gamma^{\s_{{q}},\s}_{A,\lambda_A}[p;\bar{p}] 
    \epsilon^{*\lambda_A}\left({p}+{\bar{p}}\right)\cdot J\left({p}+{\bar{p}}\right)\,.
\end{split}\end{align}
These expressions depend on the gluon formation time when it is radiated
off the $\bar q$ line,
\begin{equation}
    \bar{t}_{g} = \frac{2\bar{z}_g (1-z_q)E_\gamma}{(\k-\bar{z}_g \bar{\p})^2}\,,
\end{equation}
and on the corresponding vertex
\begin{equation}\label{eq:vertexAmpqbar}
    \bar{\gamma}^{\s, \s_{\bar{q}}}_{g,\lambda_g}\left[\bar{p};{k}\right] = 
    \bar{v}^{\s}\left(\bar p+k\right) \slashed{\epsilon}^{*\lambda_g}\left({k}\right) 
    v^{\s_{\bar{q}}}\left(\bar{p}\right)\,.
\end{equation}

\section{$\gamma^*\to q\bar q g$ matrix-element in a dense medium}
\label{sec:me}

Equipped with the $M_q$ and $M_{\bar{q}}$ amplitudes derived in the previous section, we can now calculate the squared matrix element $\mathcal{M}^2_{q\bar q g}$ for 
$\gamma^*\to q\bar q g$ splittings,
\begin{equation}
\label{eq:total-me}
\mathcal{M}^2_{q\bar q g} = \langle |M_q|^2\rangle + \langle |M_{\qbar}|^2\rangle + \langle 2 ~\re~M_q M^*_{\qbar} \rangle\,,
\end{equation}
where the product of amplitudes has also been averaged over medium configurations, and the standard average over initial and sum over final quantum numbers is implicit. 
We refer to the first two contributions in Eq.~\eqref{eq:total-me} as {\it{direct terms}}, since they correspond to the gluon being emitted by either the quark or the anti-quark both in the amplitude and the conjugate amplitude. The last contribution is referred to as the {\it{interference term}}.

\begin{figure}
    \begin{subfigure}[]{0.32\textwidth}
    \centering
    \includegraphics[width=.8\textwidth]{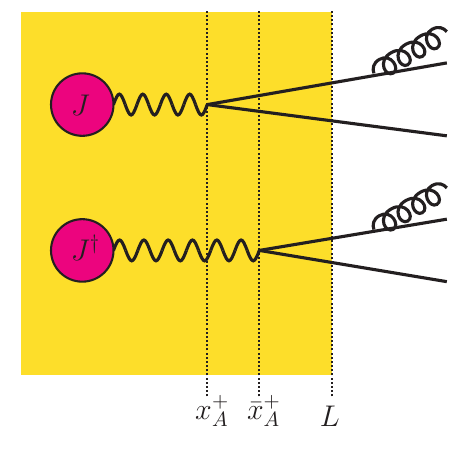}
    \caption{{\bf in-in}}
    \label{fig:m2InIn}
    \end{subfigure}
    \begin{subfigure}[]{0.32\textwidth}
    \centering
    \includegraphics[width=.8\textwidth]{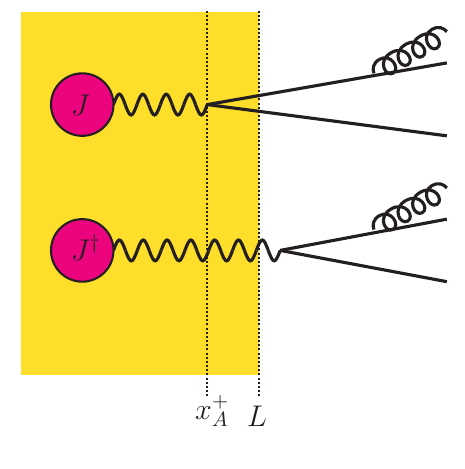}
    \caption{{\bf in-out}}
    \label{fig:m2InOut}
    \end{subfigure}
    \begin{subfigure}[]{0.32\textwidth}
    \centering
    \includegraphics[width=.8\textwidth]{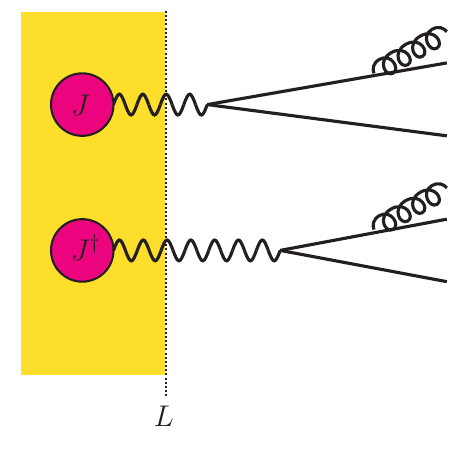}
    \caption{{\bf out-out}}
    \label{fig:m2OutOut}
    \end{subfigure}
    \caption{Different contributions to 
    Eq.~\eqref{eq:time-separation}. The amplitude
    is depicted on top of the corresponding conjugate
    amplitude. For the {\bf in-out} contributions we omit
    the diagram where the roles of the amplitude and
    conjugate amplitude are reversed.}
    \label{fig:m2}
\end{figure}

Each of the contributions in Eq.~\eqref{eq:total-me} can be written as the sum over three regions depending on the light-cone time of the antenna formation, see \cref{fig:m2}: the \textbf{in-in} region, where the antenna is created inside the medium in amplitude and conjugate amplitude; the \textbf{in-out} region, where one of the antennas is created inside and the other one outside the medium; and the \textbf{out-out} region, where both are created once the photon has escaped the medium. For instance, given the decomposition in Eq.~\eqref{eq:mq-in-out-split}, we will compute $|M_q|^2$ as the sum of
\begin{align}\begin{split} \label{eq:time-separation}
    &|M_q|^2_{\textbf{in-in}} = M^{\textbf{in}}_q M^{\textbf{in},\dagger}_q, \qquad |M_q|^2_{\textbf{in-out}} = M^{\textbf{in}}_q M^{\textbf{out},\dagger}_q + M^{\textbf{out}}_q M^{\textbf{in},\dagger}_q, \\ &|M_q|^2_{\textbf{out-out}} = M^{\textbf{out}}_q M^{\textbf{out},\dagger}_q\,.
\end{split}\end{align}

In the various terms above we need to sum and average over the corresponding quantum numbers. 
This procedure affects both the product of the polarization vectors
and the vertices given in Eqs.~\eqref{eq:verticesAmp} and \eqref{eq:vertexAmpqbar}. The former simply becomes
\begin{equation}
    \epsilon^{\lambda_A}(p+\bar p)\cdot J(p+\bar p) \, \left(\epsilon^{\bar{\lambda}_A}(p+\bar p)\cdot J(p+\bar p)\right)^\dag \;  \rightarrow  \; \d^{\lambda_A \bar{\lambda}_A}|J(p+\bar p)|^2  \, .
\end{equation}
The latter, including the average over the photon polarization 
corresponding to the factor 1/2 below, give 
\begin{align}\begin{split} \label{eq:splitting-functions}
    &\frac{1}{2} \sum \, \gamma^{\s_q,\s}_{A,\lambda_A}[p;\bar{p}] \, \bar{\gamma}^{\s, \s_{\bar{q}}}_{g,\lambda_g}[\bar{p};k] \, \left[\gamma^{\s_q,\bar{\s}}_{A,\lambda_A}[p;\bar{p}] \, \bar{\gamma}^{\bar{\s}, \s_{\bar{q}}}_{g,\lambda_g}[\bar{p}; k] \right]^* = \frac{32 E_\gamma^2 z_q}{t_{A}t_{g}z_g} P_{\gamma\to q\bar q}(z_q)\,, \\
    & \frac{1}{2} \sum \gamma^{\s_q, \s}_{g,\lambda_g}[p;k] \, \gamma^{\s,\s_{\bar{q}}}_{A,\lambda_A}[p;\bar p] \, \left[\gamma^{\s_q,\bar{\s}}_{A,\lambda_A}[p;\bar{p}] \, \bar{\gamma}^{\bar{\s}, \s_{\bar{q}}}_{g,\lambda_g}[\bar{p};k]\right]^* = \bkappa\cdot \bkappabar \frac{16 E_\gamma}{\bar z_g z_g t_{A}} P_{\gamma\to q\bar q}(z_q) \,,
\end{split}\end{align}
where $P_{\gamma\to q\bar q}(z_q) = n_f[z_q^2+(1-z_q)^2]$ is the vacuum splitting function, and we have introduced
\begin{equation}
\bkappa=\k-z_g \p, \qquad \bkappabar=\k-\bar{z}_g\pbar
\end{equation}
i.e.,~the relative transverse momentum of the gluon with respect to the quark and the anti-quark respectively.

\subsection{Direct terms}
\label{sec:dir-term}

\paragraph{In-in region:}After summing and averaging over quantum
numbers, the squared amplitude is given by 
\begin{align}
\begin{split}
    \label{eq:mq-in-in-1}
    \sum |M_q|^2_{\textbf{in-in}} &=2\, \re \, C_F\, \left( \frac{g\, e}{E_\gamma}\right)^2 \, \frac{2 t_g}{z_q t_A}\frac{1}{z_g} \, P_{\gamma\to q\bar q}(z_q) |J(p+\bar p)|^2 \, \\
    & \times \,  \int_{0}^{L} d\bar{x}^+_A \int_{0}^{\bar{x}^+_A} dx^+_A \, e^{-\ii \frac{\Delta{x}^+_A}{t_A} } \cW_{kl}\left(L,x^+_A;\r_q^A(\tau)\right) \, \cW^{\dag}_{li}\left(L,x^+_A;\r_{\bar{q}}^A(\tau)\right) \, 
    \\ & 
    \times
    \cW_{i\bar{l}}\left(L,\bar{x}^+_A;\bar{\r}_{\bar{q}}^A(\tau)\right) \, 
    \cW^{\dag}_{\bar{l} k}\left(L,\bar{x}^+_A;\bar{\r}_q^A(\tau)\right) \,,
\end{split}
\end{align}
where the trajectory of the Wilson lines in the amplitude is parametrized as in Eq.~\eqref{eq:w-line-traj-q} and in the conjugate amplitude by
\begin{equation} \begin{split}
\label{eq:w-line-traj-qbar}
   \bar{\r}_{q}^A(\tau)= \frac{\p}{E_q}(\tau-\bar{x}^+_A)\,,  \qquad\qquad \bar{\r}_{\bar{q}}^A (\tau) = \frac{\bar{\p}}{E_{\bar{q}}} (\tau-\bar{x}^+_A) \,.
\end{split}\end{equation}
We have chosen $\Delta{x}^+_A=\bar x_A^+ - x_A^+>0$, accounting for the alternative ordering by taking twice the real part of the expression. 
The Casimir of the fundamental representation
appears through
$t^{b}_{\bar{k}j} \,t^{b}_{jk} = C_F \, \d_{\bar{k} k}$. 

The next step is to average $|M_q|^2$ over medium configurations. 
Given that the medium average is local in the $+$ component (see Eq.~\eqref{eq:amu-field}), we use the composition law of Wilson lines 
\begin{equation}
\label{eq:wilson-lines-split}
\cW_{ab}\left(x^+_1,x^+_2;\r(\tau)\right) = \cW_{ac}\left(x^+_1,x^+_3;\r(\tau)\right) 
 \, \cW_{cb}\left(x^+_3,x^+_2;\r(\tau)\right)
\end{equation}
to rewrite the squared amplitude as
\begin{align} 
    \label{eq:mq-in-in-2}
   & \sum \langle|M_q|^2_{\textbf{in-in}} \rangle =2\, \re \, C_F\, \left( \frac{g\, e}{E_\gamma}\right)^2 \frac{2 t_g}{z_q t_{A}} \,\frac{1}{z_g}P_{\gamma\to q\bar q}(z_q) |J(p+\bar p)|^2 \, \nonumber 
    \\ & \times \int_{0}^{L} d\bar{x}^+_A \int_{0}^{\bar{x}^+_A} dx^+_A \, 
    e^{-\ii\frac{\Delta{x}^+_A} {t_{A}}} \Big<\cW_{ml}\left(\bar{x}^+_A,x^+_A;\r_q^A(\tau)\right) \, \cW^{\dag}_{ln}\left(\bar{x}^+_A,x^+_A;\r_{\bar{q}}^A(\tau)\right) \Big>\, 
    \\ & \times \Big< \, \cW_{km}\left(L,\bar{x}^+_A;\r_q^A(\tau)\right) \, \cW^{\dag}_{ni}\left(L,\bar{x}^+_A;\r_{\bar{q}}^A(\tau)\right) \cW_{i\bar{l}}\left(L,\bar{x}^+_A;\bar{\r}_{\bar{q}}^A(\tau)\right) \, \cW^{\dag}_{\bar{l} k}\left(L,\bar{x}^+_A;\bar{\r}_q^A(\tau)\right) \Big> \, \nonumber.
\end{align}
Then, we also use the property of colour triviality (see
Eq.~\eqref{eq:medium-correlator}) to obtain
\begin{align}\begin{split}
 \label{eq:W-trace}
    &\Big<\cW_{ml}\left(\bar{x}^+_A,x^+_A;\r_q^A(\tau)\right) \, \cW^{\dag}_{ln}\left(\bar{x}^+_A,x^+_A;\r_{\bar{q}}^A(\tau)\right) \Big> \,
    \\ &= \frac{\d_{mn}}{N_c} \Big< \Tr  \, \cW\left(\bar{x}^+_A,x^+_A;\r_q^A(\tau)\right) \, \cW^{\dag}\left(\bar{x}^+_A,x^+_A;\r_{\bar{q}}^A(\tau)\right) \Big> \, ,
\end{split}\end{align}
and similarly for the medium average of four Wilson lines. 
Putting everything together we get our final expression for
the \textbf{in-in} contribution to the direct term
\begin{align} \begin{split}
 \label{eq:mq-in-in}
    \sum \langle|M_q|^2_{\textbf{in-in}} \rangle &= 4\,C_F N_c \, \left( \frac{g\, e}{E_\gamma}\right)^2 \, \frac{t_g}{z_q t_A} \, \frac{1}{z_g} P_{\gamma\to q\bar q}(z_q) |J(p+\bar p)|^2\, 
    \\ &\times \int_{0}^{L} d\bar{x}^+_A \int_{0}^{\bar{x}^+_A} dx^+_A \, \cos\left(\frac{\Delta{x}^+_A}{t_A}\right) \mathcal{S}^{(2)}\left(\bar{x}^+_A,x^+_A; \r^A_q(\tau), \r^A_{\bar{q}}(\tau)\right) \,  
    \\ & \times \, \mathcal{Q}^{(4)}\left(L,\bar{x}^+_A; \r^A_q(\tau), \r^A_{\bar{q}}(\tau), \bar{\r}^A_q(\tau),
    \bar{\r}^A_{\bar{q}}(\tau)\right)  \, .
\end{split}\end{align}
We have introduced the two-point function $\mathcal{S}^{(2)}$, usually referred to as `dipole' in the literature,
\begin{equation}
\label{eq:dipole-def}
    \mathcal{S}^{(2)}\left(\bar{x}^+_A,x^+_A; \r^A_q(\tau) , \r^A_{\bar{q}}(\tau) \right) 
    = \frac{1}{N_c} \Big< \Tr  \, \cW\left(\bar{x}^+_A,x^+_A;\r^A_q(\tau)\right) \, \cW^{\dag}\left(\bar{x}^+_A,x^+_A;\r^A_{\bar{q}}(\tau)\right) \Big> \, ,
\end{equation}
and the four-point function $\mathcal{Q}^{(4)}$, the so-called `quadrupole',
\begin{align}
\label{eq:quadrupole-def}
    &\mathcal{Q}^{(4)}\left(L,\bar{x}^+_A; \r^A_q(\tau), \r^A_{\bar{q}}(\tau), \bar{\r}^A_q(\tau),
    \bar{\r}^A_{\bar{q}}(\tau)\right) \,
    \\ &  = \frac{1}{N_c} \Big< \Tr \, \cW\left(L,\bar{x}^+_A;\r^A_q(\tau)\right) \, \cW^{\dag}\left(L,\bar{x}^+_A; \r^A_{\bar{q}}(\tau)\right) \, \cW\left(L,\bar{x}^+_A; \bar{\r}^A_{\bar{q}}(\tau)\right) \, \cW^{\dag}\left(L,\bar{x}^+_A;\bar{\r}^A_q(\tau)\right) \Big> \, \nonumber.
\end{align}
Both the dipole and the quadrupole are purely real functions.

In the first line of Eq.~\eqref{eq:mq-in-in} we recognise the vacuum
splitting function for the gluon emission in the soft limit 
$P_{q\to qg}=1/z_g$. 
This observation renders a clear separation between the gluon 
emission process and the medium dynamics, as expected given the 
limits we have taken throughout our calculation.

\paragraph{In-out region:} We next turn to the case in which the antenna is created outside the medium either in the amplitude or in the conjugate amplitude. We find that the squared amplitude is given by
\begin{align}\begin{split}\label{eq:mq-in-out-1}
    \sum |M_q|^2_{\textbf{in-out}} & = 2\,\ii C_F \, \left( \frac{g\, e}{E_\gamma}\right)^2 \, \frac{t_g}{z_qz_g} \, P_{\gamma\to q\bar q}(z_q)|J(p+\bar p)|^2  \, 
    \\ & \times \Bigg\{
    \int_0^L d x^+_A \, e^{\ii \frac{(L-x^+_A)}{t_A}} \,  \cW_{il}\left(L,x^+_A;\r^A_{\bar{q}}(\tau)\right) \, \cW^{\dag}_{li}\left(L,x^+_A;\r^A_q(\tau)\right)
     \\ &\quad
    -\int_0^L dx^+_A \, e^{-\ii \frac{(L-x^+_A)}{t_A}} \,  \cW_{il}\left(L,x^+_A;\r^A_q(\tau)\right) \, \cW^{\dag}_{li}\left(L,x^+_A;\r^A_{\bar{q}} (\tau)\right) \, \Bigg\} \, . 
\end{split}\end{align}
Note that after performing the medium average, the two terms in the curly brackets are identical up to the sign in the phase. Therefore, we obtain
\begin{align}\begin{split}
\label{eq:mq-in-out}
    \sum \langle |M_q|^2_{\textbf{in-out}} \rangle &= -4\,C_F N_c \left( \frac{g\, e}{E_\gamma}\right)^2 \, \frac{t_{g}}{z_g z_q} \, P_{\gamma\to q\bar q}(z_q) |J(p+\bar p)|^2 \,
    \\ &  \times \int_0^L dx^+_A \, \sin\left(\frac{L-x^+_A}{t_A} \right) \, \mathcal{S}^{(2)}\left(L,x^+_A;\r^A_q(\tau),\r^A_{\bar{q}}(\tau)\right) \, ,
\end{split}\end{align}
where we have introduced the dipole as defined in Eq.~\eqref{eq:dipole-def}.

\paragraph{Out-out region:} 
Lastly, we study the case in which the antenna is formed outside the medium.
All Wilson lines become Kronecker deltas in colour space,
and all integrals can be trivially computed to get
\begin{equation}
\label{eq:mq-out-out}
    \sum \langle |M_q|^2_{\textbf{out-out}} \rangle = 2 \,C_F N_c \, \left( \frac{g\, e}{E_\gamma}\right)^2 \, \frac{t_g t_A}{z_q z_g} \, P_{\gamma\to q\bar q}(z_q)|J(p+\bar p)|^2 \,,
\end{equation}
which coincides with the direct quark contribution to the squared matrix 
element for gluon radiation from a colour singlet antenna in vacuum.

The full result for the direct quark contributions is 
obtained by summing Eqs.~\eqref{eq:mq-in-in}, \eqref{eq:mq-in-out}
and \eqref{eq:mq-out-out}. The direct anti-quark contributions
can be obtained from the quark results by replacing $z_g\to \bar z_g$, 
$z_q\to\bar{z}_q= (1-z_q)$ and $t_{g}\to \bar t_{g}$.

\subsection{Interference term}
\label{sec:int-term}
Here we compute the contribution to the squared matrix element 
in which the radiation is emitted from one leg of the antenna in 
the amplitude, and absorbed by the other one in the conjugated amplitude.
These diagrams are responsible for interference effects between the 
gluon emission and the dynamics of the antenna formation, as we shall
discuss in more detail below. 

\paragraph{In-in region:} 
We start by focusing on the region where the antenna is created 
inside the medium. The squared matrix element is given by 
\begin{align}\begin{split}
\label{eq:mqmqbar-in-in-1}
    2\,\re & \sum \left( M_q  M^*_{\bar{q}}\right)_{\textbf{in-in}}
    =
    - 2\,\re \, \frac{(ge)^2}{z_q(1-z_q) E^3_\gamma}  \, \frac{t_g \bar t_g}{\bar z_g z_g t_A} \, |J(p+\bar{p})|^2 \, \bkappa \cdot \bkappabar \, P_{\gamma\to q\bar q}(z_q) \,
    \\ & \times  \int_{0}^{L} d\bar{x}^+_A \int_{0}^{\bar{x}^+_A} dx^+_A \,\Big[ e^{-\ii \frac{\Delta{x}^+_A}{t_A}}t^{b}_{jk}\,\cW_{kl}\left(L,x^+_A;\r^A_q(\tau)\right) \, \cW^{\dag}_{li}\left(L,x^+_A;\r^A_{\bar{q}}(\tau)\right) \, 
    \\ & \times \, t^{b}_{i\bar k}\,\cW_{\bar{k}\bar l}\left(L,\bar{x}^+_A;\bar{\r}^A_{\bar{q}}(\tau)\right) \, \cW^{\dag}_{\bar{l}j}\left(L,\bar{x}^+_A;\bar{\r}^A_q(\tau)\right) \, + \left(x^+_A \leftrightarrow \bar{x}^+_A \right) \Big] ,
\end{split}\end{align}
where we accounted for the two possible orderings of the light-cone 
times of the antenna splitting. Let us comment on the colour structure. 
By using the Fierz identity we can rewrite the product of SU($N_c$) 
generators as
\begin{equation}
t^{b}_{jk} t^{b}_{i\bar k} = \frac{1}{2} \left(\d_{j\bar{k}} \, \d_{ki}-\frac{1}{N_c} \,\d_{jk} \, \d_{i\bar{k}}\right).
\end{equation}
The first term of the previous equation, when combined with 
the colour structure of the Wilson lines, leads to
\begin{align}\begin{split}
\label{eq:mqmqbar-in-in-2}
    2\,\re & \sum \left( M_q  M^*_{\bar{q}}\right)_{\textbf{in-in}}
     = - \re \, \frac{(ge)^2}{z_q(1-z_q) E^3_\gamma}  \, \frac{t_g \bar t_g}{\bar z_g z_g t_A} \, |J(p+\bar{p})|^2 \, \bkappa \cdot \bkappabar \, P_{\gamma\to q\bar q}(z_q) \,
    \\ & \times  \int_{0}^{L} d\bar{x}^+_A \int_{0}^{\bar{x}^+_A} dx^+_A \,\Big[ e^{-\ii \frac{\Delta{x}^+_A}{t_A}}\,\cW_{kl}\left(L,x^+_A;\r^a_q(\tau)\right) \, \cW^{\dag}_{lk}\left(L,x^+_A;\r^A_{\bar{q}}(\tau)\right) \, 
    \\ & \times \,\cW_{j\bar l}\left(L,\bar{x}^+_A;\bar{\r}^A_{\bar{q}}(\tau)\right) \, \cW^{\dag}_{\bar{l}j}\left(L,\bar{x}^+_A;\bar{\r}^A_q(\tau)\right) \, + \left(x^+_A \leftrightarrow \bar{x}^+_A \right) \Big] .
\end{split}\end{align}
We neglect the contribution arising from the second term in 
the Fierz identity since we work in the large-$N_c$ limit.

Averaging Eq.~\eqref{eq:mqmqbar-in-in-2} over all possible 
configurations of the background field and using their locality 
and colour triviality leads to
\begin{align}
    2&\,\re \sum \langle M_q  M^*_{\bar{q}}\rangle_{\textbf{in-in}}
    = - 4\,\re \, C_F N_c\, \frac{(ge)^2}{z_q(1-z_q) E^3_\gamma}  \, \frac{t_g \bar t_g}{\bar z_g z_g t_A} \, |J(p+\bar{p})|^2 \bkappa \cdot \bkappabar \, P_{\gamma\to q\bar q}(z_q) \, \nonumber
    \\ & \times \, \int_{0}^{L} d\bar{x}^+_A \int_{0}^{\bar{x}^+_A} dx^+_A \Big[ \, e^{-\ii \frac{\Delta{x}^+_A}{t_A}}  \frac{1}{N_c} \Big< \cW_{ml}\left(\bar{x}^+_A,x^+_A;\r^A_q(\tau)\right) \, \cW^{\dag}_{lm}\left(\bar{x}^+_A,x^+_A;\r^A_{\bar{q}}(\tau) \right)\Big> \label{eq:mqmqbar-in-in-3}
    \\ & \times \frac{1}{N_c^2} \,\Big< \cW_{in}\left(L,\bar{x}^+_A;\r^A_q(\tau)\right) \, \cW^{\dag}_{ni}\left(L,\bar{x}^+_A;\r^A_{\bar{q}}(\tau)\right) \, \cW_{j\bar{l}}\left(L,\bar{x}^+_A;\bar{\r}^A_{\bar{q}}(\tau)\right) \, \cW^{\dag}_{\bar{l}j}\left(L,\bar{x}^+_A;\bar{\r}^A_q(\tau)\right) \Big>\Big], \nonumber
\end{align}
where the overall factor of 4 comes from writing 
$N_c^2=2C_FN_c$ in the large-$N_c$ limit and 
taking into account the two orderings of the integration
variables written explicitly in Eq.~\eqref{eq:mqmqbar-in-in-2}.
We identify two structures of Wilson-line 
correlators: a dipole between $x_A^+$ and $\bar x_A^+$, 
$\mathcal{S}^{(2)}\left(\bar{x}^+_A, x^+_A; \r^A_q(\tau), 
\r^A_{\bar{q}}(\tau)\right)$ (see Eq.~\eqref{eq:dipole-def}), 
and the product of four Wilson lines that are traced into 
two pairs, the so-called `double-dipole'. 
In the large-$N_c$ limit, it can be shown that 
this object reduces to the product of two dipoles
(see e.g.~Ref.~\cite{Isaksen:2020npj}).
Therefore, we can further simplify Eq.~\eqref{eq:mqmqbar-in-in-3} 
and obtain 
\begin{align}
\label{eq:mqmqbar-in-in}
     2&\,\re \sum \langle M_q  M^*_{\bar{q}}\rangle_{\textbf{in-in}} = -  4\, C_F N_c\, \frac{(ge)^2}{z_q(1-z_q) E^3_\gamma}  \, \frac{t_g \bar t_g}{\bar z_g z_g t_A} \, |J(p+\bar{p})|^2 \bkappa \cdot \bkappabar \, P_{\gamma\to q\bar q}(z_q)\, \nonumber
    %
    \\ & \times  \int_{0}^{L} d\bar{x}^+_A \int_{0}^{\bar{x}^+_A} dx^+_A \,\cos \left(\frac{\Delta{x}^+_A}{t_A}\right)\mathcal{S}^{(2)}\left(\bar{x}^+_A, x^+_A; \r^A_q(\tau), \r^A_{\bar{q}}(\tau)\right) 
    \\ & \times \, \mathcal{S}^{(2)}\left(L,\bar{x}^+_A; \r^A_q(\tau), \r^A_{\bar{q}}(\tau)\right)\mathcal{S}^{(2)}\left(L,\bar{x}^+_A; \bar{\r}^A_{\bar{q}}(\tau), \bar{\r}^A_q(\tau)\right) \nonumber\, . 
\end{align}

\paragraph{In-out region:} The colour structure heavily simplifies in this situation since, analogously to the direct term calculation, we have to deal with only two Wilson lines. The result is
\begin{align}\begin{split}
\label{eq:mqmqbar-in-out-1}
     2\,\re &\sum \left( M_q  M^*_{\bar{q}}\right)_{\textbf{in-out}} = 2\,\ii\re \, C_F \, \frac{(ge)^2}{z_q(1-z_q) E^3_\gamma}  \, \frac{t_g \bar t_g}{\bar z_g z_g} \, |J(p+\bar{p})|^2 \, \bkappa \cdot \bkappabar \, P_{\gamma\to q\bar q}(z_q)\, 
    \\ & \times \Bigg\{\int_0^L dx^+_A \, e^{-\ii \frac{(L-x^+_A)}{t_A}} \,  \cW_{il}\left(L,x^+_A;\r^A_q(\tau)\right) \, \cW^{\dag}_{li}\left(L,x^+_A;\r^A_{\bar{q}} (\tau)\right) \, 
    \\ & \hspace{0.3cm} - \int_0^L dx^+_A \, e^{\ii \frac{(L-x^+_A)}{t_{A}}} \,  \cW_{li}\left(L,x^+_A;\r^A_{\bar{q}} (\tau)\right) \, \cW^{\dag}_{il}\left(L,x^+_A;\r^A_q(\tau)\right) \Bigg\} \, .
\end{split}\end{align}
After performing the medium average, this expression reduces to 
\begin{align}\begin{split}
\label{eq:mqmqbar-in-out}
    2\,\re \sum \langle M_q  M^*_{\bar{q}}\rangle_{\textbf{in-out}}& =  4\, C_FN_c \, \frac{(ge)^2}{z_q(1-z_q) E^3_\gamma}  \, \frac{t_{g} \bar t_{g} }{\bar z_g z_g}\, |J(p+\bar{p})|^2 \bkappa \cdot \bkappabar\, P_{\gamma\to q\bar q}(z_q) \, 
    \\ & \times \int_0^L dx^+_A \, \sin\left(\frac{L-x^+_A}{t_A}\right) \, \mathcal{S}^{(2)}\left(L,x^+_A;\r^A_q(\tau),\r^A_{\bar{q}} (\tau)\right) \, ,
\end{split}\end{align}
where we identify the same functional form for the medium contribution as the one present in the direct term, see Eq.~\eqref{eq:mq-in-out}.

\paragraph{Out-out region:} Setting the antenna to be outside the medium
in both contributions leads to
\begin{equation}
\label{eq:mqmqbar-out-out}
     2\,\re \sum \langle M_q  M^*_{\bar{q}}\rangle_{\textbf{out-out}} = - 2 \, C_F N_c \,\frac{(g e)^2}{z_q(1-z_q) E^3_\gamma} \, \frac{t_{g} \bar t_{g}  t_{A}}{\bar z_g z_g} \, |J(p+\bar{p})|^2 \, \bkappa \cdot \bkappabar \, P_{\gamma\to q\bar q}(z_q) \, ,
\end{equation}
which coincides with the vacuum result for the interference term.

\subsection{Medium Modifications to Squared Matrix Elements}
\label{sec:final-res}

The full result for the direct terms, including the emission off 
the quark line and that off the anti-quark, can be compactly written as 
\begin{equation}
\label{eq:direct-term}
    \langle|M_q|^2 \rangle + \langle|M_{\bar q}|^2 \rangle = 
     4\,C_F g^2 \mathcal{M}^{2}_{q\bar q}\left(\frac{1}{\bkappa^2}+\frac{1}{\bkappabar^2}\right) \,
   \left(1+\Fmed\right)\,,
\end{equation}
where we have introduced $\mathcal{M}^{2}_{q\bar q}$,
which corresponds to the production of a $q\bar q$ antenna from a virtual
photon in vacuum, and is given by
\begin{equation}
\label{eq:qqbarVac}
    \mathcal{M}^{2}_{q\bar q}
    =\frac{e^2 N_c t_A}{E_\gamma}\, P_{\gamma\to q\bar q}(z_q) 
    |J(p+\bar p)|^2\,.
\end{equation}
We have identified $F_{\rm med}$, the known 
factor controlling the medium-induced modification to the 
$q\bar q$ antenna production \cite{Dominguez:2019ges}
\begin{align}\begin{split}\label{eq:fmedDef}
    \Fmed & = 2 \int_{0}^{L} \frac{d\bar{x}^+_A}{t_A} \int_{0}^{\bar{x}^+_A} \frac{dx^+_A}{t_A} \, \cos\left(\frac{\Delta{x}^+_A}{t_A}\right) \mathcal{S}^{(2)}\left(\bar{x}^+_A,x^+_A; \r^A_q(\tau), \r^A_{\bar{q}}(\tau)\right) \,  
    \\ & \times \, \mathcal{Q}^{(4)}\left(L,\bar{x}^+_A; \r^A_q(\tau), \r^A_{\bar{q}}(\tau), 
    \bar{\r}^A_q(\tau),
    \bar{\r}^A_{\bar{q}}(\tau)
    \right)   \,   \\
   & -2 \int_0^L \frac{dx^+_A}{t_A} \, \sin\left(\frac{L-x^+_A}{t_A} \right) \, 
   \mathcal{S}^{(2)}\left(L,x^+_A;\r^A_q(\tau),\r^A_{\bar{q}}(\tau)\right)\,.
\end{split}\end{align}
We will discuss the limiting behaviour of this expression
in the following section.

By summing Eqs.~\eqref{eq:mqmqbar-in-in}, \eqref{eq:mqmqbar-in-out} and \eqref{eq:mqmqbar-out-out} we obtain the interference term, which reads
\begin{align}
\label{eq:int-term}
   & \langle 2\,\re \, M_q M^*_{\bar q}\rangle = 
    -8\,C_F g^2 \mathcal{M}^{2}_{q\bar q}\frac{\bkappa \cdot \bkappabar}{\bkappa^2 \bkappabar^2} \,\nonumber
\Big[1+ 2\int_{0}^{L} \frac{d\bar{x}^+_A}{t_A} \int_{0}^{\bar{x}^+_A} \frac{dx^+_A}{t_A} \,\cos \left(\frac{\Delta{x}^+_A}{t_A}\right)\nonumber
    \\ & \times \, \mathcal{S}^{(2)}\left(L,\bar{x}^+_A; \r^A_q(\tau), \r^A_{\bar{q}}(\tau)\right) \mathcal{S}^{(2)}\left(L,\bar{x}^+_A; \bar{\r}^A_{\bar{q}}(\tau), \bar{\r}^A_q(\tau)\right) \mathcal{S}^{(2)}\left(\bar{x}^+_A, x^+_A; \r^A_q(\tau), \r^A_{\bar{q}}(\tau)\right) \\
    & - 2 \, \int_0^L \frac{dx^+_A}{t_A} \, \sin\left(\frac{L-x^+_A}{t_A}\right) \, \mathcal{S}^{(2)}\left(L,x^+_A;\r^A_q(\tau),\r^A_{\bar{q}} (\tau)\right) \Big] \,  \nonumber .
\end{align}
We then recast this expression as
\begin{equation}\label{eq:int-term-alt}
    \langle 2\,\re \, M_q M^*_{\bar q}\rangle =
    -8\,C_F g^2 \mathcal{M}^{2}_{q\bar q}\frac{\bkappa \cdot \bkappabar}{\bkappa^2 \bkappabar^2}(1+\Fmed) \,
    \left(1- \TildeDeltamed\right)
\end{equation}
where we have defined $\TildeDeltamed$ as
\begin{align}\begin{split}
\label{eq:tilde-delta-med}
\TildeDeltamed &= \frac{2}{1+\Fmed} \int_{0}^{L} \frac{d\bar{x}^+_A}{t_A} \int_{0}^{\bar{x}^+_A} \frac{dx^+_A}{t_A} \, \cos\left(\frac{\Delta{x}^+_A}{t_A}\right) \mathcal{S}^{(2)}\left(\bar{x}^+_A,x^+_A; \r^A_q(\tau), \r^A_{\bar{q}}(\tau)\right) \,   
    \\ & \times \, \Big[\mathcal{Q}^{(4)}\left(L,\bar{x}^+_A; \r^A_q(\tau), \r^A_{\bar{q}}(\tau), 
    \bar{\r}^A_q(\tau),
    \bar{\r}^A_{\bar{q}}(\tau)\right) \\
    &\quad  -\mathcal{S}^{(2)}\left(L,\bar{x}^+_A; \r^A_q(\tau), \r^A_{\bar{q}}(\tau)\right)
    \mathcal{S}^{(2)}\left(L,\bar{x}^+_A; \bar{\r}^A_{\bar{q}}(\tau), \bar{\r}^A_q(\tau)\right)\Big]. 
\end{split}\end{align}
In the rest of this paper we will refer to $\TildeDeltamed$ 
as the \emph{generalized  decoherence factor}. Indeed, as we
will see in the next section, in the limit of $\Delta x_A^+= \bar{x}^+_A - x^+_A\to 0$ it reduces to the $\Deltamed$
computed in Refs.~\cite{Mehtar-Tani:2010ebp,Mehtar-Tani:2011hma,Casalderrey-Solana:2011ule,Mehtar-Tani:2011vlz,Mehtar-Tani:2011lic,Armesto:2011ir,Mehtar-Tani:2012mfa,Casalderrey-Solana:2012evi,Calvo:2014cba}. 

We can interpret this formula as follows. First, the propagation of the virtual antenna during $\Delta x_A^+$ is described by a dipole correlator of Wilson lines. From $\bar x_A^+$ to $L$, the antenna is now real and there are two possible ways of connecting the colour of the four Wilson lines~\cite{Blaizot:2012fh}. One, described by the quadrupole, in which the $q$ in the amplitude is colour connected to the $\bar q$ in the amplitude and similarly for the $\bar q$. The second colour configuration, described by the double-dipole, consists in considering the four-particle system as two separate $q\bar q$ dipoles, one in the amplitude and the other in the conjugated amplitude.
 
Putting direct contributions and interference
terms together, we obtain our final result for the squared 
matrix element for the emission of a soft gluon off a 
(colour singlet) $q\bar q$ antenna,
\begin{equation}
\label{eq:cool-result}
    \mathcal{M}^2_{q\bar q g} =
    4\,C_F g^2 \mathcal{M}^{2}_{q\bar q}(1+\Fmed)\left[\left(\frac{1}{\bkappa^2}+\frac{1}{\bkappabar^2}
    -2
    \frac{\bkappa \cdot \bkappabar}{\bkappa^2 \bkappabar^2}\right) +2
    \frac{\bkappa \cdot \bkappabar}{\bkappa^2 \bkappabar^2} \TildeDeltamed \right].
\end{equation}
This is the main result of this paper. 
It contains two pieces: (i) the full vacuum spectrum 
multiplied by the nuclear modification factor $(1+\Fmed)$, 
and (ii) a pure interference contribution multiplied by the 
generalized decoherence factor, which depends on the medium properties and the kinematics of the antenna.

\subsection{Limiting behaviour}
We now discuss how to recover two important limits of 
Eq.~\eqref{eq:cool-result}, namely the vacuum baseline and 
the limit where we set $\Delta x_A^+$ to zero.

In the vacuum limit, all dipoles and quadrupoles in 
Eqs.~\eqref{eq:fmedDef} and \eqref{eq:tilde-delta-med}
become unity, and it follows that $\Fmed$ and $\TildeDeltamed$
vanish. 
Eq.~\eqref{eq:cool-result} then reproduces the well known
squared matrix element for soft gluon radiation off a colour 
singlet dipole in vacuum, 
see e.g.~Refs.~\cite{Dokshitzer:1991wu,Ellis:1996mzs}.

Let us now turn to the limit where $\Delta x^+_A=0$.
This is the setup considered
in Refs.~\cite{Mehtar-Tani:2010ebp,Mehtar-Tani:2011hma,Casalderrey-Solana:2011ule,Mehtar-Tani:2011vlz,Mehtar-Tani:2011lic,Armesto:2011ir,Mehtar-Tani:2012mfa,Casalderrey-Solana:2012evi,Calvo:2014cba}, 
where the antenna forms at $x^+_A=0$ (and hence 
$\bar x^+_A=0$) and the decoherence factor was found to be
\begin{equation}
\label{eq:standard-delta-med}
\Deltamed=1-\mathcal{S}^{(2),\text{adj}}\left(L,0; \r^A_q(\tau), \r^A_{\bar{q}}(\tau)\right)\,.
\end{equation}
Recovering this limiting behaviour from Eq.~\eqref{eq:cool-result}
requires some care, as some of the steps in our calculation
are only valid if the antenna forms at different times
in the amplitude and complex-conjugate amplitude. 
In particular, we note that out of the \textbf{in-in},
\textbf{in-out} and \textbf{out-out} regions only
the \textbf{in-in} region survives. Furthermore, 
when computing the \textbf{in-in} contributions
we fixed $\bar x_A^+>x_A^+$ and accounted for the
other ordering by multiplying the expressions by a factor
of 2 (see e.g.~Eqs.~\eqref{eq:mq-in-in-1} and 
\eqref{eq:mqmqbar-in-in-2}). This factor must be removed
if the time at which the antenna is formed is the same
in the amplitude and complex-conjugate amplitude.
Finally, to fix the time of formation we insert
$t_A\delta(x_A^+)$ and $t_A\delta(\bar{x}^+_A)$
in our expressions for the \textbf{in-in} contributions.
Following this procedure, we find that
the direct contributions become
\begin{equation}
    \langle|M_q|^2 \rangle + \langle|M_{\bar q}|^2 \rangle
    \to
    4\,C_F g^2 \mathcal{M}^{2}_{q\bar q}\left(\frac{1}{\bkappa^2}+\frac{1}{\bkappabar^2}\right)\,,
\end{equation}
where we used that 
\begin{equation}\label{eq:limitsDipAndQuad}
    \mathcal{S}^{(2)}\left(0,0; \r^A_q(\tau), \r^A_{\bar{q}}(\tau)\right) =1\,,\quad
    \mathcal{Q}^{(4)}\left(L,0; \r^A_q(\tau), \r^A_{\bar{q}}(\tau), 
    \r^A_q(\tau),
    \r^A_{\bar{q}}(\tau)
    \right)=1\,.
\end{equation}
As expected, we recover the fact that
in this limit there are no medium modifications to the
direct terms. 
Effectively, this amounts to taking $1+\Fmed\to 1$.
To compute the limit of the interference contribution,
we set $1+\Fmed\to 1$ and then apply the limiting
procedure described above to the integrals in $\TildeDeltamed$.
Using Eq.~\eqref{eq:limitsDipAndQuad} and noting that,
in the large-$N_c$ limit,
\begin{equation}
    \left[\mathcal{S}^{(2)}\left(L,0; \r^A_q(\tau), \r^A_{\bar{q}}(\tau)\right)\right]^2  = \mathcal{S}^{(2),\text{adj}}\left(L,0; \r^A_q(\tau), \r^A_{\bar{q}}(\tau)\right)\,,
\end{equation}
we find that
\begin{equation}
    \TildeDeltamed\to 
    \Deltamed\,,
\end{equation}
with $\Deltamed$ as defined in 
Eq.~\eqref{eq:standard-delta-med}. 
We finally get
\begin{equation}
    \langle 2\,\re \, M_q M^*_{\bar q}\rangle \to
    -8\,C_F g^2 \mathcal{M}^{2}_{q\bar q}\frac{\bkappa \cdot \bkappabar}{\bkappa^2 \bkappabar^2}
    \left(1- \Deltamed\right)\,,
\end{equation}
in full agreement with the results of Refs.~\cite{Mehtar-Tani:2010ebp,Mehtar-Tani:2011hma,Casalderrey-Solana:2011ule,Mehtar-Tani:2011vlz,Mehtar-Tani:2011lic,Armesto:2011ir,Mehtar-Tani:2012mfa,Casalderrey-Solana:2012evi,Calvo:2014cba}.

\section{Medium model and numerical results}
\label{sec:numerics}

\subsection{Medium model}
To numerically evaluate the expressions presented in the previous section, we need to specify a medium model. Following standard practice in the jet-quenching literature, we opt for taking the harmonic oscillator approximation for the in-medium potential. We further consider a static medium, i.e., a medium with constant transport properties along different trajectories,
modeled by the parameter $\hat q$. 
Note that the transport coefficient depends on the colour representation, and we choose $\hat q$ to be that corresponding to the fundamental degrees of freedom. The adjoint case is simply given by $C_A\hat q/C_F$, which in the large-$N_c$ limit reduces to $2\hat q$. Furthermore, it is convenient to work in the frame where the virtual photon has zero transverse momentum (see e.g.~the discussion in Ref.~\cite{Dominguez:2019ges}).

Under the harmonic approximation, the dipole correlator of two Wilson
lines becomes~\cite{Blaizot:2012fh,Apolinario:2014csa,Sievert:2019cwq}
\begin{equation}
\label{eq:dip-ho-aprox}
\mathcal{S}^{(2)}\left(t,t_1; \r_1 , \r_2 \right) 
    = e^{-\frac{1}{4}\hat q\int_{t_1}^{t} d\tau [\r_1(\tau) - \r_2(\tau)]^2} \, ,
\end{equation}
where $\r_{1,2}$ correspond to Wilson-line trajectories (either in the amplitude or the conjugate amplitude) as defined in Eqs.~\eqref{eq:w-line-traj-q} and \eqref{eq:w-line-traj-qbar}. 
In the large-$N_c$ limit, the quadrupole can be written in terms of these dipoles as \cite{Blaizot:2012fh,Apolinario:2014csa,Sievert:2019cwq} 
\begin{align}\label{eq:quad-gen}
    &\mathcal{Q}^{(4)}\left(L,\bar{x}^+_A; \r^A_q(\tau), \r^A_{\bar{q}}(\tau), \bar{\r}^A_q(\tau),
    \bar{\r}^A_{\bar{q}}(\tau)\right) = \mathcal{S}^{(2)}\left(L,\bar{x}^+_A; \r^A_q , \bar{\r}^A_q \right) 
    \, \mathcal{S}^{(2)}\left(L,\bar{x}^+_A; \r^A_{\bar{q}} , \bar{\r}^A_{\bar{q}} \right) \,
    \\ & \!+ \int_{\bar{x}^+_A}^L ds \, \mathcal{S}^{(2)}\left(L, s; \r^A_q , \bar{\r}^A_q \right) \mathcal{S}^{(2)}\left(L, s; \r^A_{\bar{q}} , \bar{\r}^A_{\bar{q}} \right)  T(s) \mathcal{S}^{(2)}\left(s,\bar{x}^+_A; \r^A_q , \r^A_{\bar{q}} \right) \mathcal{S}^{(2)}\left(s,\bar{x}^+_A; \bar{\r}^A_{q} , \bar{\r}^A_{\bar{q}} \right) \nonumber,
\end{align}
where 
\begin{equation}
T(s)= -\frac{\hat{q}}{2}\left[\left(\r^A_q - \r^A_{\bar q}\right)^2 + \left(\bar{\r}^A_q - \bar{\r}^A_{\bar q}\right)^2 - \left(\r^A_q - \bar{\r}^A_{\bar q}\right)^2  - \left(\bar{\r}^A_q - \r^A_{\bar q}\right)^2 \right]\, ,
\end{equation}
is the well-known transition amplitude.

These expressions can be made more explicit by inserting the trajectories for the Wilson lines. As an example, for the dipole 
$\mathcal{S}^{(2)}\left(L,x^+_A; \r^A_q , \r^A_{\bar {q}} \right)$,
the difference of transverse coordinates that appears is given by
\begin{equation}
(\r^A_q - \r^A_{\bar q})^2 = (\tau-x_A^+)^2\theta^2_{q\bar q}\,,\\ 
\end{equation}
and therefore the dipole becomes
\begin{equation}
\mathcal{S}^{(2)}\left(L,x^+_A; \r^A_q , \r^A_{\bar {q}} \right)
=e^{-\frac{1}{12}\hat q\theta^2_{q\bar q}(L-x^+_A)^3}\,.
\end{equation}
Another correlator of Wilson lines that enters Eq.~\eqref{eq:tilde-delta-med} is the double-dipole, which under the aforementioned limits and approximations reads~\cite{Isaksen:2020npj}
\begin{equation}
\label{eq:double-dip-ho-aprox}
\mathcal{S}^{(2)}\left(L,\bar{x}^+_A; \r^A_q , \r^A_{\bar {q}} \right) \mathcal{S}^{(2)}\left(L,\bar{x}^+_A; \bar{\r}^A_q , \bar{\r}^A_{\bar {q}} \right) = e^{-\frac{1}{12}\hat q\theta^2_{q\bar q}[(L-\bar{x}^+_A)^3-(\bar{x}^+_A-x_A^+)^3+(L-x^+_A)^3]}\, .
\end{equation}
Finally, starting from Eq.~\eqref{eq:quad-gen}, the quadrupole entering Eqs.~\eqref{eq:fmedDef} and \eqref{eq:tilde-delta-med} is \cite{Blaizot:2012fh,Apolinario:2014csa,Sievert:2019cwq}
\begin{align}
\label{eq:quad-ho-aprox}
&\mathcal{Q}^{(4)}\left(L,\bar{x}^+_A; \r^A_q, \r^A_{\bar{q}}, 
    \bar{\r}^A_q,
    \bar{\r}^A_{\bar{q}}\right)
     = e^{-\frac{1}{4}\hat q \theta^2_{q\bar q}[z_q^2+(1-z_q)^2](\bar{x}_A^+-x_A^+)^2(L-\bar{x}_A^+)} \\
    \nonumber
    & + T(\bar{x}_A^+, x_A^+) \int_{\bar{x}_A^+}^L ds \,e^{-\frac{1}{4}\hat q\theta_{q\bar q}^2[z_q^2+(1-z_q)^2](L-s)(\bar{x}_A^+- x_A^+)^2} \, e^{-\frac{1}{12}\hat q\theta_{q\bar q}^2[(s-\bar{x}_A^+)^3+(s-x_A^+)^3-(L-s)^3]}\,,
\end{align}
with the transition amplitude becoming
\begin{equation}
\label{eq:transition-amplitude}
T(\bar{x}_A^+, x_A^+)= -\hat q z_q(1-z_q) \theta_{q\bar q}^2 (\bar{x}_A^+ -x_A^+)^2.
\end{equation}
In the previous equations we have introduced the opening angle of the antenna, $\theta_{q\bar q}=|\boldsymbol{\theta}_{q\bar q}|$, that is defined as the modulus of
\begin{equation}
\label{eq:thqqbar-def}
\boldsymbol{\theta}_{q\bar q}\equiv\frac{\p}{z_q E_\gamma} - \frac{\pbar}{ (1-z_q)E_\gamma}\,.
\end{equation}
We note that everything in the latter expression is defined in light-cone coordinates. The light-cone energies and angles are related to the usual energy $p^0$ and Cartesian angle $\theta^{\rm cart}_{q\bar q}$ by a rescaling (see footnote \ref{foot:LCvars}), $E=p^+\sim \sqrt{2}p^0$ and $\theta_{q\bar q} = \theta^{\rm cart}_{q\bar q} /\sqrt{2}$. 
Following this convention, the antenna formation time $t_A$ defined in Eq.~\eqref{eq:formation-times} becomes
\begin{equation}
    t_A=\frac{2}{z_q(1-z_q)\theta_{q\qbar}^2E_\gamma}\,.
\end{equation}

Within the specific medium model we have chosen, 
the various correlators of Wilson lines discussed above
are governed by characteristic physical scales. As an example,
we note that $\Deltamed$ can be written as
\begin{equation}
    \Deltamed=1-e^{-\frac{1}{6}\hat q\theta^2_{q\bar q}L^3}
    =1-e^{-\theta^2_{q\bar q}/\theta^2_{c}}\,,
\end{equation}
where we have introduced the well-known critical 
angle~\cite{Mehtar-Tani:2006vpj}
\begin{equation}
\label{eq:thetac-def}
\theta_c = \sqrt{\frac{6}{\hat q L^3}}.
\end{equation}
Other relevant scales can be identified in both
$\Fmed$ and $\TildeDeltamed$. These have already been
investigated in detail in Refs.~\cite{Dominguez:2019ges,Isaksen:2023nlr} within the context of the calculation of $\Fmed$ so we do not discuss them again here. 

\subsection{Numerical results and discussion}

\begin{figure}
      \includegraphics[width=0.32\textwidth]{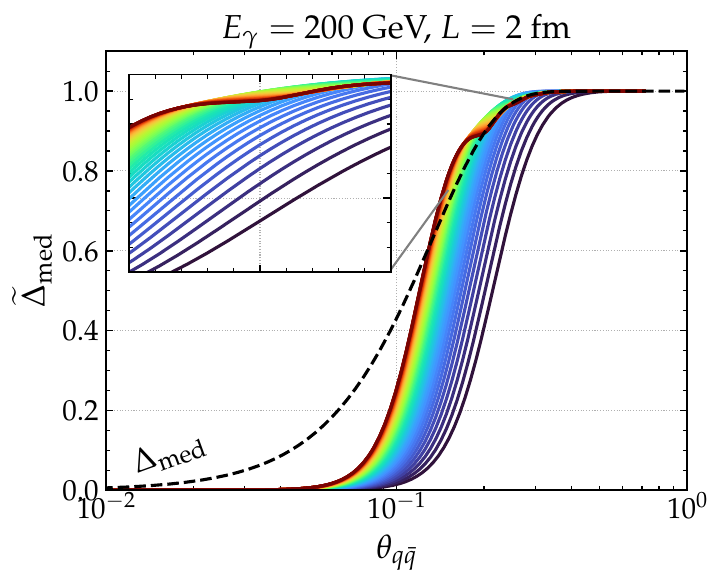}
       \includegraphics[width=0.32\textwidth]{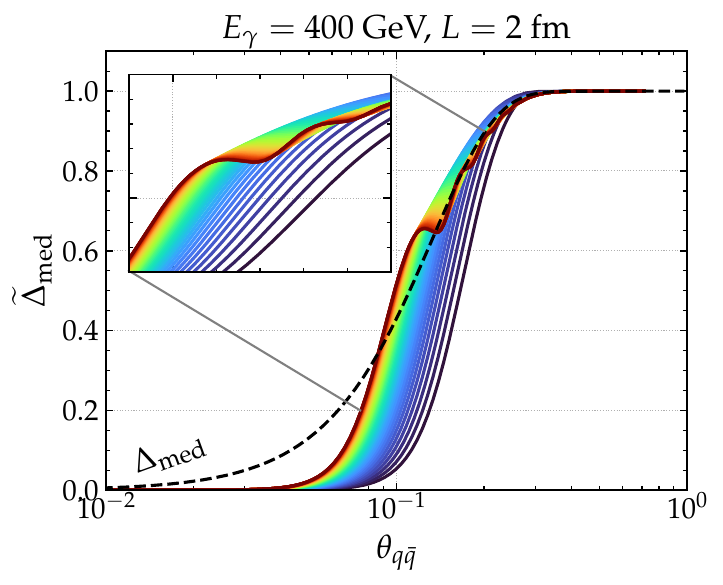}
       \includegraphics[width=0.32\textwidth]{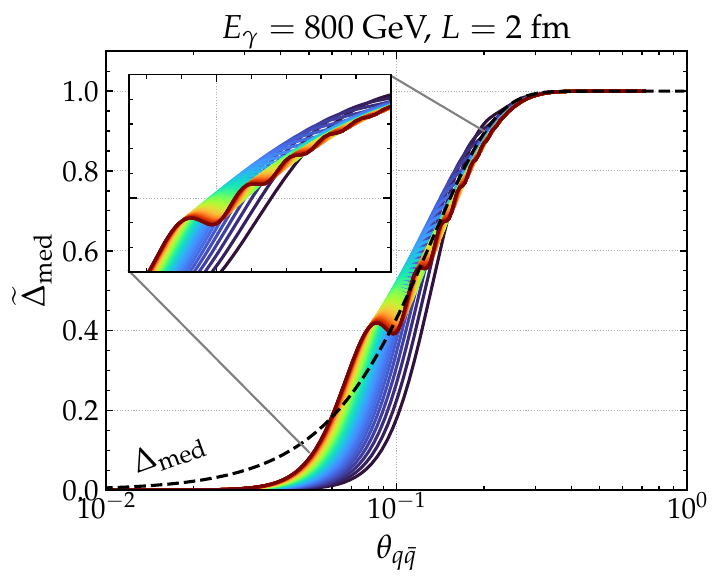}
       \\
        \includegraphics[width=0.32\textwidth]{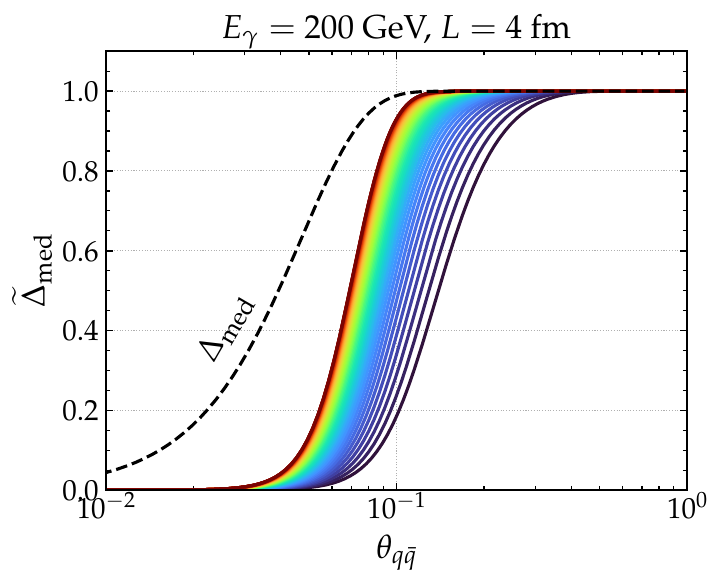}
       \includegraphics[width=0.32\textwidth]{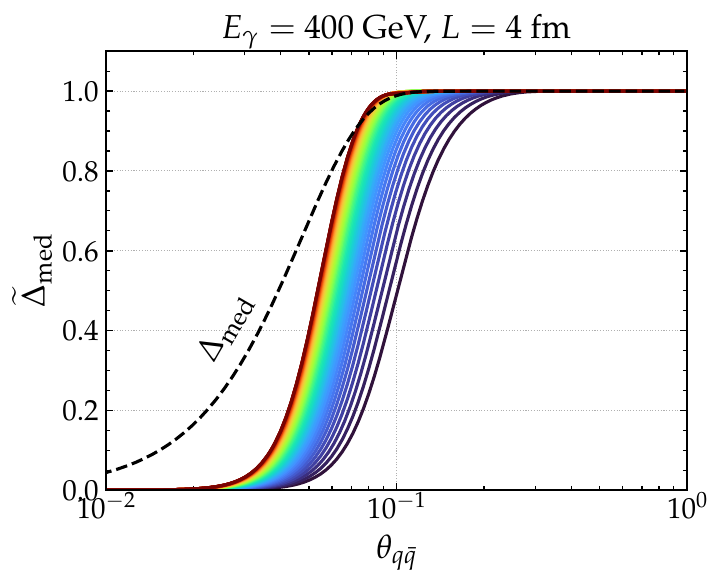}
       \includegraphics[width=0.32\textwidth]{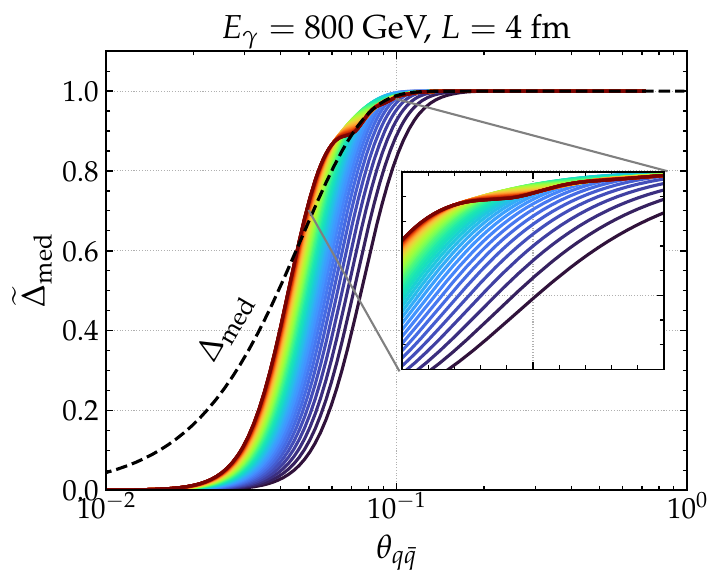}
       \\
      \includegraphics[width=0.32\textwidth]{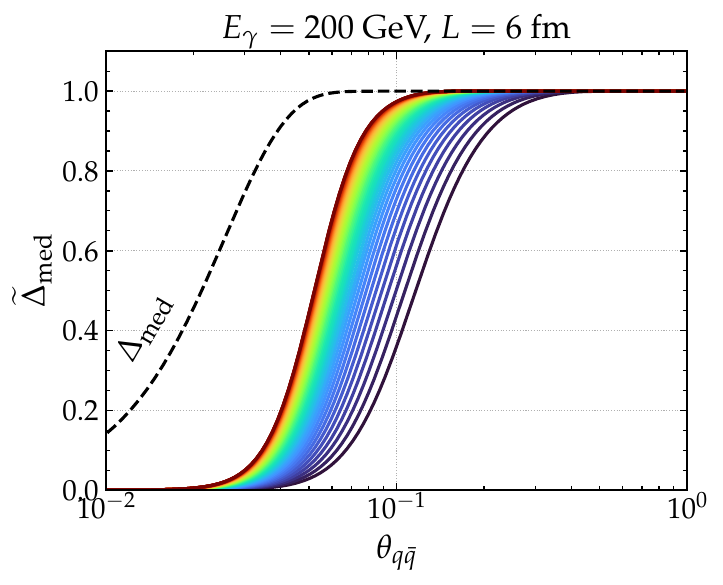}
      \includegraphics[width=0.32\textwidth]{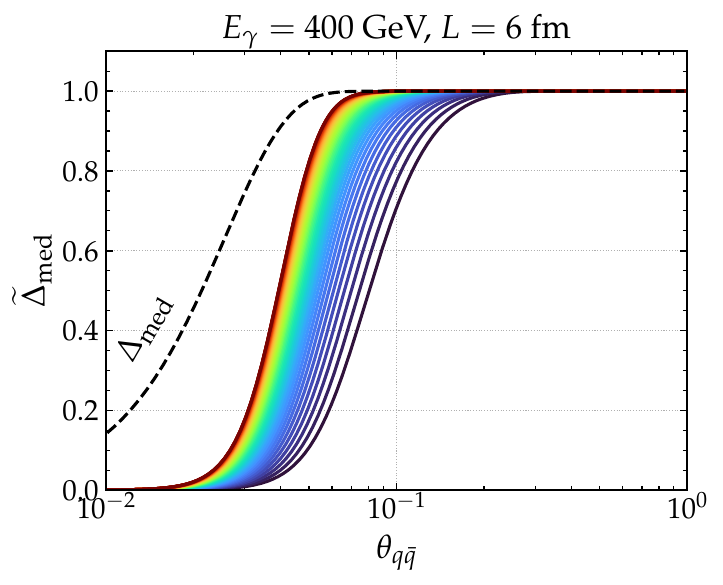}
    \includegraphics[width=0.32\textwidth]{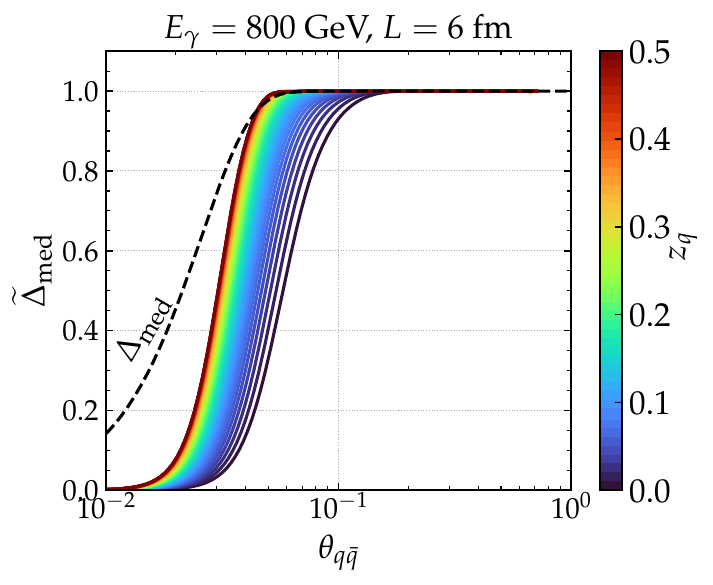}
    \caption{Numerical evaluation of the generalized decoherence factor $\TildeDeltamed$ in Eq.~\eqref{eq:tilde-delta-med} as a function of the opening angle of the antenna, $\theta_{q\bar q}$, and its energy sharing fraction, $z_q$, for three different photon energies and medium lengths. The dashed black line corresponds to $\Deltamed$, see Eq.~\eqref{eq:standard-delta-med}. The value of $\hat q$ is fixed to $1.5$ GeV$^2$/fm.}
    \label{fig:plot1}
\end{figure}

In this section we present a numerical study of our main
result, $\TildeDeltamed(E_\gamma,z_q,\theta_{q\bar q})$,
as defined in Eq.~\eqref{eq:tilde-delta-med}, within
the medium model discussed above. 
Together with this manuscript, we include an ancillary
file, \texttt{tildeDeltaMedNum.m}, that contains a simple numerical implementation
of $\TildeDeltamed$ (and $\Fmed$, see Eq.~\eqref{eq:fmedDef}) in 
\texttt{Mathematica} that should
allow the reader to reproduce our results in the bulk
of the parameter space we explore. The numerical
results discussed below were obtained with a more robust
numerical implementation.

We first evaluate the generalized decoherence factor $\TildeDeltamed(E_\gamma,z_q,\theta_{q\bar q})$ and compare it to the standard $\Deltamed(\theta_{q\bar q})$ for a medium with fixed $\hat q=1.5$~GeV$^2$/fm, varying lengths $L=(2,4,6)$~fm and with photon energies $E_{\gamma}=(200,400,800)$ GeV.\footnote{Note that we take these values for the light-cone coordinates used throughout this paper, see the comment below Eq.~\eqref{eq:thqqbar-def}.} These medium parameters were chosen to be consistent with other jet-quenching phenomenological studies using the static brick approximation, see e.g.~Refs.~\cite{Caucal:2019uvr,Mehtar-Tani:2021fud,Andres:2022ovj,Attems:2022otp}. Our results are presented in Fig.~\ref{fig:plot1}.

Before delving into the behavior of $\TildeDeltamed$, let us briefly comment on the standard $\Deltamed$ result. This function only depends on the medium properties, and, as such, it remains invariant when increasing $E_\gamma$ (moving from left to right in Fig.~\ref{fig:plot1}). Its parametric dependence on the medium is also transparent: given that $\hat q$ is fixed, the larger the 
$L$, the smaller the value of $\theta_c$  (see Eq.~\eqref{eq:thetac-def}). 
In other words, for larger $L$ we get full decoherence of
the gluon radiation at smaller antenna opening angles.

We start the description of $\TildeDeltamed$ by focusing on the  collinear regime ($\theta_{q\bar q}\ll 1$). We observe that $\TildeDeltamed$ is systematically bigger than $\Deltamed$ for all values of $z_q$. This observation can be understood analytically by expanding Eqs.~\eqref{eq:cool-result} and \eqref{eq:standard-delta-med} in the small-angle limit. We find that they behave as 
\begin{align}\begin{split}
\label{eq:small-angle-expansion}
&\TildeDeltamed\Big\vert_{\theta_{q\bar q}\ll 1} = 
\frac{1}{480}\hat q L^5E_\gamma^2z_q^2(1-z_q)^2(3-2z_q+2z_q^2)\theta_{q\bar q}^6
+\mathcal{O}(\theta_{q\bar q}^8)\\
&\Deltamed\Big\vert_{\theta_{q\bar q}\ll 1} =\frac{1}{6}
\hat qL^3\theta_{q\bar q}^2+\mathcal{O}(\theta_{q\bar q}^4)
\end{split}\end{align}
We note that the radius of convergence of these expansions is very small: while for $\theta_{q\bar q}<0.01$ they give a good approximation of the exact result, for $\theta_{q\bar q}>0.01$ the series stops converging very quickly for realistic values of $E_\gamma$ and $\hat{q}$.
Nevertheless, the small-angle expansions provide an explanation for the behaviour we observe in all the plots of \cref{fig:plot1}, i.e., $\TildeDeltamed<\Deltamed$ in the $\theta_{q\bar q}\to 0$ limit. Physically, this implies that gluons radiated off very hard-collinear prongs of a $q\bar q$-antenna are vacuum-like (since $\Fmed$ also vanishes in this limit, as shown in Fig.~\ref{fig:Fmeds}).

In the opposite regime, $\theta_{q\bar q}\to 1$, both decoherence factors $\TildeDeltamed$ and $\Deltamed$ reach unity. Consequently, the medium-induced correction to the antenna emission pattern exactly cancels the vacuum interference term, and thus the radiation pattern of the antenna becomes that of two independent colour charges. 
For the parameter space we explore, this regime is reached for small values of $\theta_{q\bar q}$, well before $\theta_{q\bar q}= 1$, where the small angle approximation we have taken in our calculation is still valid. 
We then conclude that the regime of total colour decoherence ($\TildeDeltamed=1$) is reached even when incorporating medium modifications during the formation of the antenna. 
However, the value of $\theta_{q\bar q}$ for which $\TildeDeltamed\to1$ is now a function not only of the medium properties but also of $z_q$ and $E_\gamma$. 

Understanding the large-angle ($\theta_{q\bar q}\to 1$) behaviour of both
$\Fmed$ and $\TildeDeltamed$ analytically is not trivial given the integrals appearing in their definition. Instead, we have numerically explored this limit, observing that
$\Fmed\to 0$ for sufficiently large values of $z_q$ (see Fig.~\ref{fig:Fmeds} and the discussion in Appendix~\ref{app:bonus}). We have also probed the large-angle limit of each of the two contributions appearing in the numerator of $\TildeDeltamed$ numerically.
We found that terms proportional to the quadrupole go to $1$, while those proportional to the double dipole go to $1/2$, reproducing the expected behaviour, 
$\TildeDeltamed\to1$.

Perhaps one of the most interesting and novel features of Fig.~\ref{fig:plot1} is the transition region at intermediate angles, where the colour decoherence factor is neither $0$ nor $1$.
For the values of the parameters explored in this work, we identify two regimes that we proceed to discuss separately. 
First, for dense enough media ($\hat q L=6, 9$ GeV$^2$, with $\hat q=1.5$ GeV$^2/$fm) and photon energies of $E_\gamma=200, 400$ GeV we find that $\TildeDeltamed$ monotonically grows with $z_q$. 
In other words, for fixed $\theta_{q\bar q}$, radiation with larger values of $z_q$ will
have a larger proportion of anti-angular ordered emissions.\footnote{An important remark is that very asymmetric splittings, when $z_q\sim 0$ (or $z_q\sim 1$), are beyond the regime of validity of our calculation since, in these cases, one of the two prongs is very soft and thus the tilted Wilson line approximation for the propagators is not justified.}
In addition, even at $z_q=0.5$, there is a gap between the $\Deltamed$ and $\TildeDeltamed$ curves, whose size varies with both $L$ and $E_\gamma$. 
This gap, more pronounced for $L=6$ fm and $E_\gamma=200, 400$ GeV, leads to a clear delay of the onset of total decoherence.
Finally, we observe that in this regime $\Deltamed$ and $\TildeDeltamed$ are not just shifted but also have different slopes. When fixing $\hat q$ and $L$ so as to fix $\Deltamed$, we find that $\TildeDeltamed$ gets closer to $\Deltamed$ when increasing $E_\gamma$. We interpret this observation as a consequence of the reduction of the antenna formation time with increasing $E_\gamma$, since then the impact of considering $\Delta x_A^+\neq 0$ is suppressed, i.e.,~the antenna formation is less sensitive to medium effects and $\TildeDeltamed\to \Deltamed$. 

When reducing the medium length to $L=2$ fm (and also for $L=4$ fm, $E_\gamma=800$ GeV), we observe new interesting features. First, a non-monotonic dependence of $\TildeDeltamed$ on $z_q$. As clearly shown in the inset plots, curves for moderate values of $z_q$ can surpass that corresponding to the democratic branching case at relatively small angles, $\theta_{q\bar q}\sim 0.1$. The insets also showcase an oscillatory behaviour of  $\TildeDeltamed$ that becomes more pronounced when increasing the photon energy. 
We would like to highlight that these oscillations are already present in $\Fmed$ when evaluated in this kinematic regime (see Fig.~\ref{fig:Fmeds}). Indeed, they result from the interplay between the frequency of the trigonometric functions 
(which is independent of the medium parameters $\hat q$ and $L$, but depends on $E_\gamma$ and $\theta_{q\bar{q}}$ through $t_A$), and the exponential decay of the Wilson-line correlators 
(that depend solely on the medium parameters).\footnote{\label{fn:oscillations} We have performed numerical checks to verify that the oscillations are not due to numerical instabilities. More precisely, we have used various numerical integration strategies (as implemented in \texttt{Mathematica}) and checked that they give the same results, and we have studied the error associated with the numerical integration to be sure that it is well below the amplitude of the observed oscillations. Furthermore, we note that the oscillations appear in regions of small $\theta_{q\bar q}$ where the numerical integrations converge well. For some choices of medium parameters, as $\theta_{q\bar q}\to 1$ the convergence becomes problematic and we have excluded those points since then we always have that $\TildeDeltamed \to 1$. } Quantitatively, the most important observation is that the gap between $\Deltamed$ and $\TildeDeltamed$ significantly shrinks, especially for the highest value of $E_\gamma$. In other words, $\Deltamed$ becomes a good approximation for $\TildeDeltamed$ despite its non-trivial parametric dependences.   

\begin{figure}
    \centering
   \includegraphics[width=0.49\textwidth]{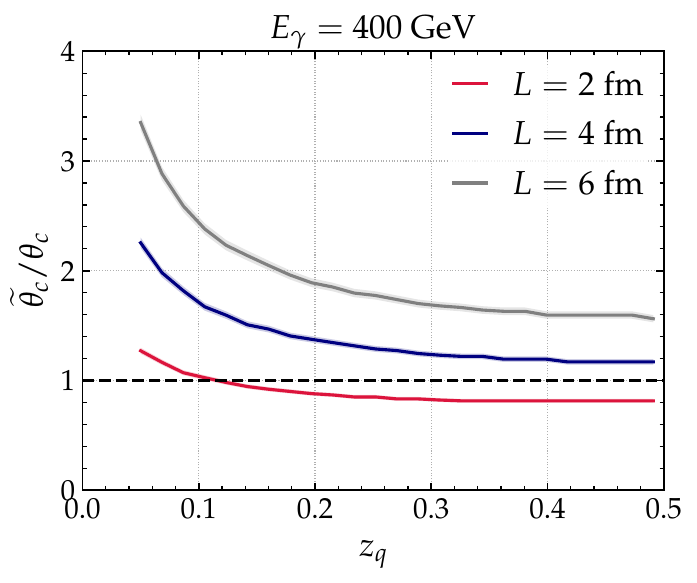}
    \includegraphics[width=0.49\textwidth]{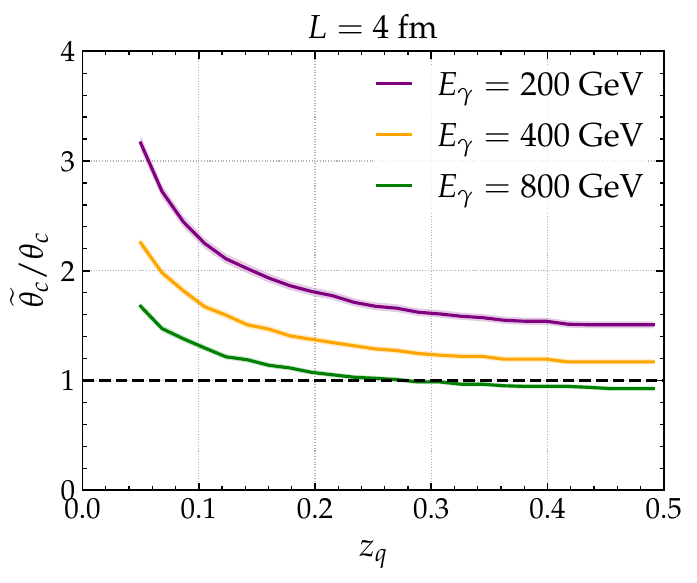}
    \caption{Ratio between the generalized critical angle and the standard one (see Eq.~\eqref{eq:thetac-def}) as a function of the energy sharing fraction for different medium lengths and photon energies.}
    \label{fig:thetacs}
\end{figure}

A way of further characterizing the colour decoherence factor is by identifying 
a generalisation of the critical angle defined in Eq.~\eqref{eq:thetac-def}.
In analogy with $\theta_c$, we define $\tilde\theta_c$ as the angle for which 
$\TildeDeltamed(E_\gamma,z_q,\theta_{q\bar q}=\tilde\theta_c)=1-e^{-1}$.
Traditionally, this angle has been interpreted as a property of the medium that controls its resolution power: splittings with $\theta_{q\bar q} < \theta_c$ were said not to be resolved by the medium. After accounting for finite formation time effects, this picture becomes more complex since this angle is not only a property of the medium but also depends on $E_\gamma$ and $z_q$, and thus changes on a splitting-by-splitting basis.

The extracted values of $\tilde\theta_c/\theta_c$ are presented in Fig.~\ref{fig:thetacs}.
This ratio allows us to quantify the aforementioned gap between $\Deltamed$ and $\TildeDeltamed$. In the left panel, the energy of the photon is fixed and the medium properties vary. In this way, both $\theta_c$ and $\tilde\theta_c$ change. 
The ratio between these two quantities shows an enhancement due to finite formation time effects for small values of $z_q$, and then a flattening as $z_q\to0.5$. 
However, since our results are not valid for $z_q\to0$, we should treat the small $z_q$ behaviour with care. 
The difference between $\theta_c$ and $\tilde\theta_c$ can be substantial.
For instance, focusing on $z_q\geq 0.2$, and on long media ($L=6$ fm), we find that the new critical angle is larger by a factor of order $2$. 
The case $L=2$ fm is again qualitatively different from the other two since $\tilde\theta_c/\theta_c$ becomes slightly smaller than one. It is important to note that, for this medium setup, the $\TildeDeltamed$ curves oscillate with a non-negligible amplitude, and thus the meaning of $\tilde\theta_c$ becomes less transparent. Indeed, 
given the oscillatory nature of the curves, we can encounter situations where 
$\TildeDeltamed=1-e^{-1}$ for two different angles.  The right panel of Fig.~\ref{fig:thetacs} shows the dependence of $\tilde\theta_c/\theta_c$ on the energy of the photon for fixed values of $\hat q$ and $L$. In this case, the denominator of this ratio is fixed and only the numerator varies. We observe that the ratio becomes smaller with increasing $E_\gamma$. This is again compatible with the gap between $\TildeDeltamed$ and $\Deltamed$ becoming narrower for shorter formation times of the antenna.

\begin{figure}
    \centering
   \includegraphics[width=0.49\textwidth]{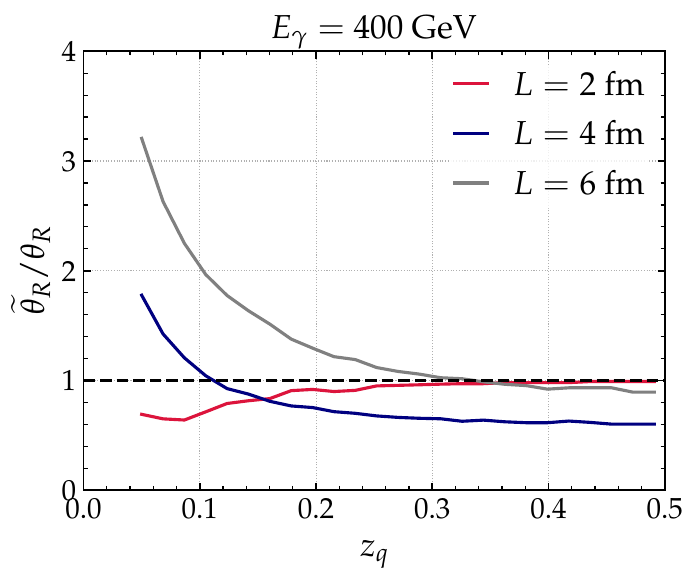}
    \includegraphics[width=0.49\textwidth]{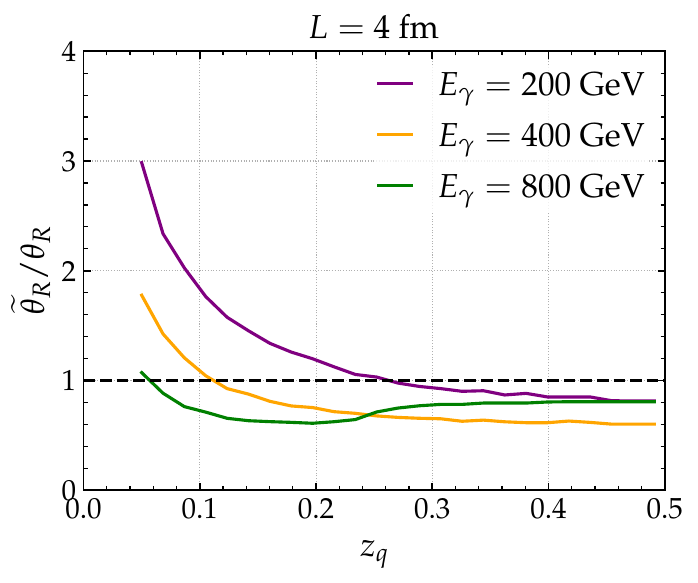}
    \caption{Ratio between the rise angle extracted for $\TildeDeltamed$ and $\Deltamed$ as a function of the energy sharing fraction for different medium lengths and photon energies.}
    \label{fig:thetars}
\end{figure}

The characterization of the differences between $\TildeDeltamed$ and $\Deltamed$
in terms of a single critical angle is somehow incomplete.
Indeed, as we have mentioned earlier, $\TildeDeltamed$ and $\Deltamed$ are not related 
by a simple shift (see Fig.~\ref{fig:plot1}). 
Furthermore, given the more involved dependence of $\TildeDeltamed$ on $\theta_{q\bar q}$, a smaller value of $\tilde\theta_c$ does not imply that $\TildeDeltamed$ reaches the regime of total colour decoherence faster.
For these reasons, we introduce another variable that we call `the rise angle', $\theta_R$. 
This variable measures the angular interval in which the colour decoherence factor rises from a given value ${\Delta}_{\rm med}^{\rm min}$ to ${\Delta}_{\rm med}^{\rm max}$ (and similarly for $\widetilde{\Delta}_{\rm med}$), informing us about the slope of the distribution:
the smaller $\theta_R$ is, the faster the distribution rises. 
We perform our extraction of $\theta_R$ in the ${\Delta}_{\rm med}^{\rm min} = 0.2$ and ${\Delta}_{\rm med}^{\rm max} = 0.95$ window so as to capture the transition region. 
As we did for $\theta_c$, we present the ratio between the value for the rise angle using $\TildeDeltamed$, denoted $\tilde\theta_R$, and the one obtained from $\Deltamed$, denoted $\theta_R$.

The results are displayed in Fig.~\ref{fig:thetars} in the same format as for Fig.~\ref{fig:thetacs}: in the left panel we fix the photon energy and change the medium parameters, and in the right panel we do the opposite. Let us again focus on the region of semi-hard splittings. The first thing to notice on the left panel is that the curve for $L=2$ fm is compatible with one. 
This supports the idea that $\Deltamed$ is a good approximation of $\TildeDeltamed$ for this regime. 
For $L=4$ fm, we find that $\TildeDeltamed$ rises almost twice as fast as $\Deltamed$ ($\tilde\theta_R/\theta_R\sim 0.5$), and $\TildeDeltamed$ is closer to a Heaviside function in this particular scenario.  
The picture is reversed when further increasing the value of $L$, i.e.,~$\tilde\theta_R/\theta_R$ is bigger or equal to 1 for $L=6$ fm. 
We see that the rise angle provides complementary information compared to that
extracted from the critical angle.
While from the critical angle perspective it seems that, 
for $L=4$ fm, $\TildeDeltamed$ 
 behaves similarly to $\Deltamed$, we observe that the rise
angle manifests the much sharper rise of $\TildeDeltamed$. Conversely, the
rise angle does not allow to capture the gap between $\TildeDeltamed$ and 
$\Deltamed$ that is clearly present for $ L=6$ fm, but this is neatly
exposed by the critical angle.

The right panel of Fig.~\ref{fig:thetars} keeps $\theta_R$ fixed. Interestingly, we observe that $\tilde\theta_R/\theta_R<1$ for sufficiently large $z_q$ values, that is $\TildeDeltamed$ rises faster than $\Deltamed$. This information could not have been extracted from the $\theta_c$ analysis, since from that we could only tell that the critical angle was larger. Putting together both observations, we conclude that even if $\tilde\theta_c>\theta_c$, the generalized decoherence factor grows faster towards the total decoherence regime.

\section{Conclusions}
\label{sec:conclusions}

One of the most striking modifications induced by the QGP on a QCD parton shower is the breaking of angular-ordering. This was observed in a series of pioneering papers almost 15 years ago by studying the radiation pattern off a QCD antenna in the presence of a medium~\cite{Mehtar-Tani:2010ebp,Mehtar-Tani:2011hma,Casalderrey-Solana:2011ule,Mehtar-Tani:2011vlz,Mehtar-Tani:2011lic,Armesto:2011ir,Mehtar-Tani:2012mfa,Casalderrey-Solana:2012evi,Calvo:2014cba}. The parameter that emerged from these studies was the so-called critical angle $\theta_c$, a property of the medium that dictates whether it is able to resolve a certain splitting or not. In this way, splittings with $\theta<\theta_c$ are not resolved and mostly respect angular ordering for subsequent emissions. Conversely, splittings with $\theta>\theta_c$ are resolved, and the radiation pattern of the antenna becomes that of two independent colour charges. The physical picture is that, for resolved splittings, interactions with the medium induce rotations of the colour state of each prong of the antenna, and they effectively become decorrelated sources in colour space. Constraining the values of $\theta_c$, and more generally the physics of colour decoherence, is one of the targets of the heavy-ion experimental program at the LHC~\cite{CMS:2017qlm,ATLAS:2022vii,ALargeIonColliderExperiment:2021mqf,ATLAS:2023hso}.    

In this work, we have revisited the calculation of 
Refs.~\cite{Mehtar-Tani:2010ebp,Mehtar-Tani:2011hma,Casalderrey-Solana:2011ule,Mehtar-Tani:2011vlz,Mehtar-Tani:2011lic,Armesto:2011ir,Mehtar-Tani:2012mfa,Casalderrey-Solana:2012evi,Calvo:2014cba}, but accounted for medium effects during the formation of the antenna. While we have restricted our calculation to a colour singlet antenna radiating a soft gluon outside of the medium, we already observed some new striking effects. In particular, we have shown that the notion of a critical angle that exclusively depends on the medium properties no longer holds after considering that the antenna itself can interact with the medium while being formed. We have introduced a generalized decoherence factor, $\TildeDeltamed$, that depends on both the medium properties ($\hat q, L$) and the kinematics of the antenna ($E_\gamma, z_q, \theta_{q\bar q}$). Consequently, each splitting in the parton shower experiences colour decoherence in a different fashion. More concretely, when considering a dense medium ($\hat q = 1.5$ GeV$^2/$fm, $L=6$ fm) we find that $\TildeDeltamed$ remains zero for a broader interval of $\theta_{q\bar q}$ values than $\Deltamed$. The direct implication is that these very collinear splittings do not violate angular ordering, i.e., their radiation behaves like vacuum emissions. In addition, we observe that there are at least two ways to reduce the differences between $\TildeDeltamed$ and $\Deltamed$: increasing the energy of the photon or reducing $L$. The interpretation of the former is very clear, as it is equivalent to reducing the antenna formation time and thus it is natural that medium modifications effects to the antenna formation are less pronounced.

Given the fact that $\TildeDeltamed$ and $\Deltamed$ are not simply related by a shift, even at fixed $z_q$, characterizing their differences in terms of differences of critical angles is insufficient. We have introduced a new metric, the rise angle $\theta_R$, that measures how fast the decoherence factor grows between two set values (we take them to be $0.2$ and $0.95$). The rise angle becomes smaller the faster the function grows, becoming zero for a Heaviside function. Within the explored region of phase-space, we find a non-trivial parametric dependence of $\tilde\theta_R$, the rise angle associated with $\TildeDeltamed$.
More precisely, we find that it can be larger or smaller than the equivalent interval for $\Deltamed$ depending on the medium and antenna properties. This indicates that accounting for medium modifications during the antenna formation can either delay or accelerate colour decoherence.

Besides the interference effects, we find that the total rate of emissions off the antenna is enhanced by a factor $\Fmed$, that was first introduced in Refs~\cite{Dominguez:2019ges,Isaksen:2023nlr}. This function also has a non-trivial parametric dependence and can be as large as a factor of $\mathcal{O}(10)$ for non-extreme values of the parameter space, e.g.~$E_\gamma=200$ GeV, $L=6$ fm, $z_q=0.2$ and $\theta_{q\bar q}= 0.15$.
Thus, medium modifications not only alter the balance between angular ordered and anti-angular ordered emissions but also induce more radiation overall.   

This generalized picture of colour decoherence has multiple consequences. As an important example, this class of interference effects is crucial when formulating an in-medium parton shower. Several implementations of colour decoherence exist in the literature~\cite{Caucal:2018dla,Hulcher:2017cpt,JETSCAPE:2023hqn}. The one closest in spirit to the dynamics encapsulated in $\Deltamed$ is that of the JetMed parton shower where the first vacuum-like emission outside of the medium can happen at any angle while angular ordering is preserved for all subsequent emissions~\cite{Caucal:2018dla}. In light of the new functional dependence of $\TildeDeltamed$, the idea of always breaking angular ordering for the first emission outside of the medium should be revisited. Phenomenologically, colour decoherence affects observables such as the fragmentation function~\cite{Caucal:2020xad} and, more generally, the jet substructure~\cite{Mehtar-Tani:2017web,Casalderrey-Solana:2019ubu,Caucal:2021cfb,Mehtar-Tani:2021fud,Pablos:2022mrx,Cunqueiro:2023vxl,Andres:2022ovj, Mehtar-Tani:2024jtd}. All these calculations use $\theta_c$ as an angular cutoff to determine whether the two prongs in a splitting lose energy by emitting medium-induced emissions. Once again, the fact that $\tilde\theta_c$ has a more intricate parametric dependence will impact these results. Furthermore, one should also study the impact of the increase in the rate of gluon emissions related to $\Fmed$ in in-medium parton showers.

In order to make the generalized picture of colour decoherence applicable to realistic LHC scenarios there are a few aspects of the calculation that should be improved. First, we need to extend the calculation to other splitting channels relevant for jet production. 
Second, it would also be interesting to explore the case in which the emission takes place inside the medium. Steps in this direction have been taken in Ref.~\cite{Barata:2021byj}.  In that work, the authors studied coherence effects for $q\to q g_1 g_2$ splittings with the gluon $g_1$ emitted inside the medium and the gluon $g_2$ outside. The more complex colour algebra of this splitting channel hampered the possibility of identifying in a clean fashion the corresponding $\Fmed$ and $\TildeDeltamed$ factors, as we were able to do in Eq.~\eqref{eq:cool-result} for a colour singlet antenna. 
An approximation that could help to simplify the calculation would be to consider that the two splittings have short formation times.
Aside from the generalization to other splittings, it would also be interesting to explore the consequences of relaxing the eikonal approximation for the prongs of the antenna and the soft limit for the emitted gluon, so that we have a better control of these effects in all regions of phase space.

Another extension of this work concerns heavy quarks. In this case, colour decoherence competes with dead-cone effects that manifest as a suppression of radiation below the dead-cone angle, $\theta_0\sim m_Q/E_Q$. Calculating the corresponding $\TildeDeltamed$ for a heavy-quark antenna would then impact recent phenomenological proposals to measure the filling of the dead-cone in heavy-ion collisions~\cite{Cunqueiro:2022svx,Andres:2023ymw}. This is so because the possibility of filling the dead-cone with medium-induced emissions requires that those emissions have angles satisfying $\theta_c < \theta < \theta_0$ region~\cite{Armesto:2003jh}.
In the standard picture, this hierarchy can be achieved by exploiting the fact that the critical angle $\theta_c$ only depends on the properties of the medium, while the dead-cone angle depends on the energy and the mass of the emitter. Given that $\tilde\theta_c$ now also depends on the kinematics of the emitter, the fine-tuning of angular scales becomes more delicate.  

Yet another direction would be to relate our work to the different perspective on the `antenna laboratory' that has been recently presented in Ref.~\cite{Pablos:2024muu}. This work studies the impact of colour coherence effects on the recoiling parton from the QGP, thus shifting the focus from the high-energy probe. We plan to carry out a dedicated study on how to incorporate these coherence effects in our setup. 
 
Finally, we note that there have been several efforts in the past to understand the angular structure of the in-medium QCD cascade beyond the antenna setups discussed above. For example, the $1\to 3$ in-medium splitting functions in the collinear limit were computed in Ref.~\cite{Fickinger:2013xwa} at first order in opacity. In this limit, vacuum/medium radiation is neither angular nor anti-angular ordered. This calculation was also revisited in Ref.~\cite{Casalderrey-Solana:2015bww} imposing additional constraints on the energy ordering of the outgoing particles. The very challenging task of computing the matrix element for two medium-induced emissions with generic kinematics is an on-going effort~\cite{Arnold:2015qya,Arnold:2022mby,Arnold:2023qwi}.
It would be interesting to understand in which limit the expressions presented in those works reduce to $\TildeDeltamed$. 

In summary, the results we have presented show that the breaking of angular ordering due to medium interactions has a more intricate behavior than previously known. 
While in this paper this was demonstrated in a simple setting, we look forward to extending our investigations to other phenomenologically relevant processes.

\section*{Acknowledgements}
We are grateful to João Barata, Paul Caucal, Fabio Dominguez, Johannes Isaksen, Andrey Sadofyev, Carlos Salgado and Konrad Tywoniuk for insightful discussions during the development of this project. 
XML and GM thank the CERN theoretical department for the hospitality at different stages of this work. The work of XML and GM is supported by European Research Council project ERC-2018-ADG-835105 YoctoLHC. XML contribution to this work is also supported  Xunta de Galicia (Centro singular de investigación de Galicia accreditation 2019-2022), by European Union ERDF; and by Grant CEX2023- 001318-M funded by MICIU/AEI/10.13039/501100011033 and by ERDF/EU, and under scholarship No. PRE2021-097748, funded by MCIN/AEI/10.13039/501100011033 and FSE+.
ASO is supported by the Ramón y Cajal program under grant RYC2022-037846-I.

\appendix
\section{Complementary material}
\label{app:bonus}

\begin{figure}
    \includegraphics[width=0.32\textwidth]{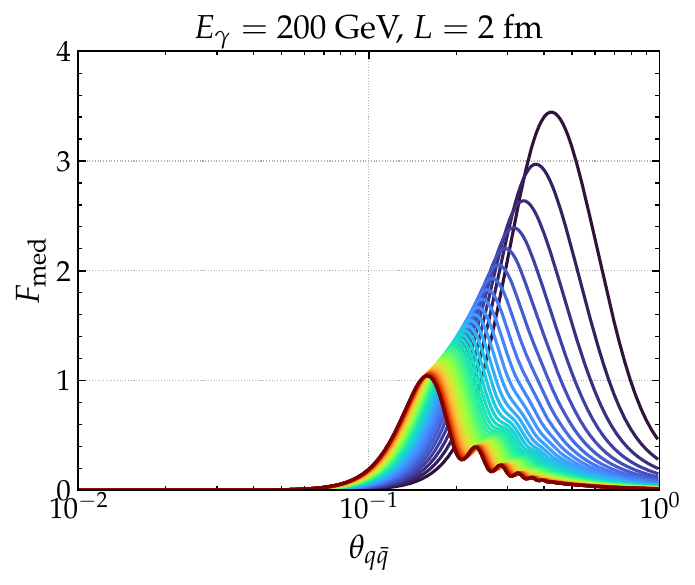}
      \includegraphics[width=0.32\textwidth]{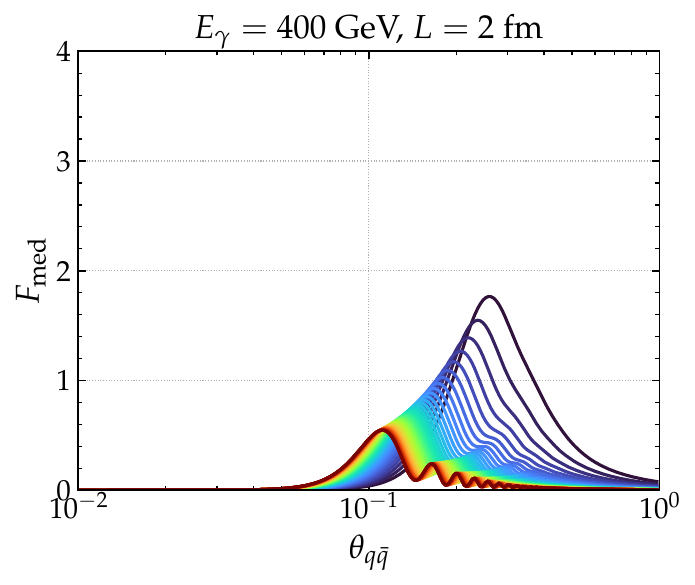}
      \includegraphics[width=0.32\textwidth]{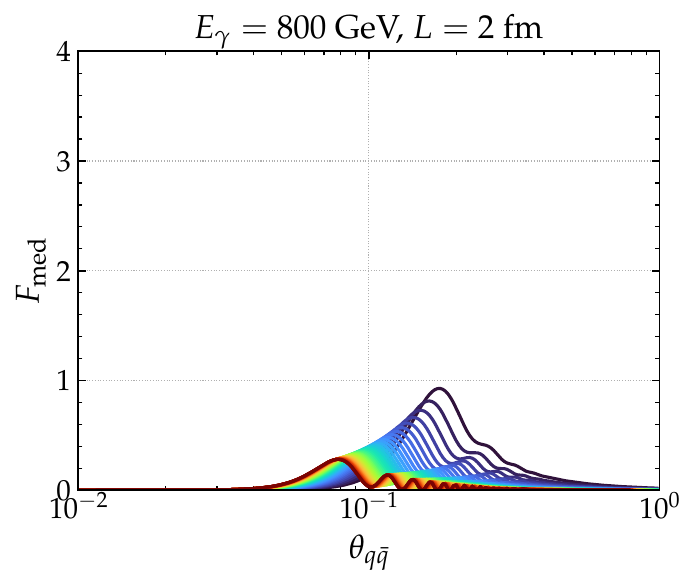}\\
      \includegraphics[width=0.32\textwidth]{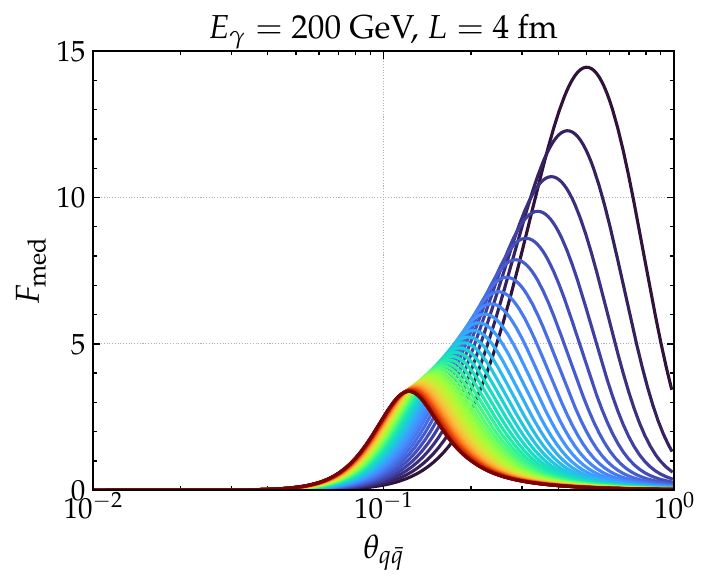}
      \includegraphics[width=0.32\textwidth]{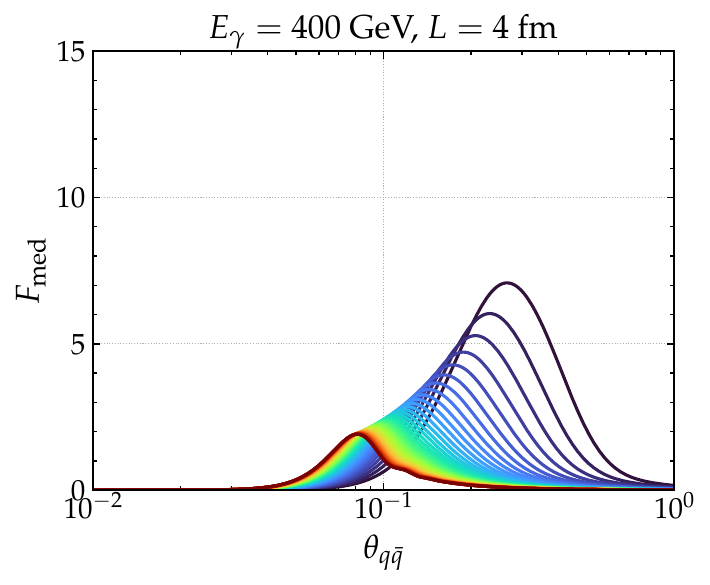}
      \includegraphics[width=0.32\textwidth]{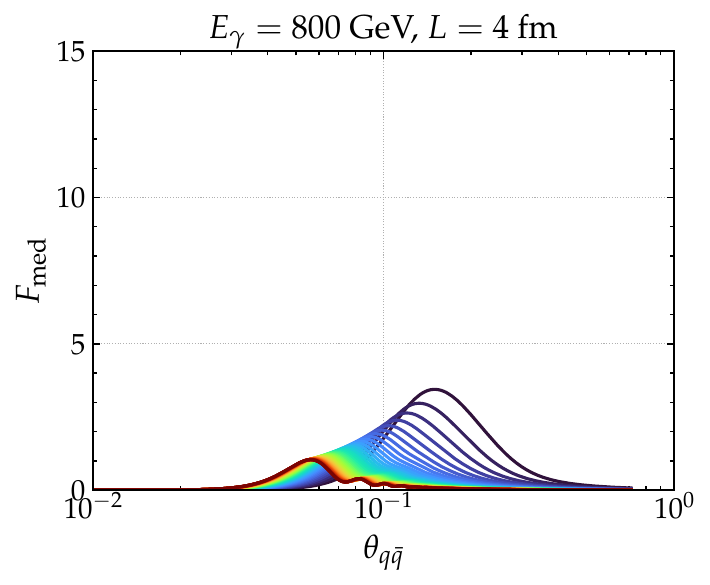} \\
       \includegraphics[width=0.32\textwidth]{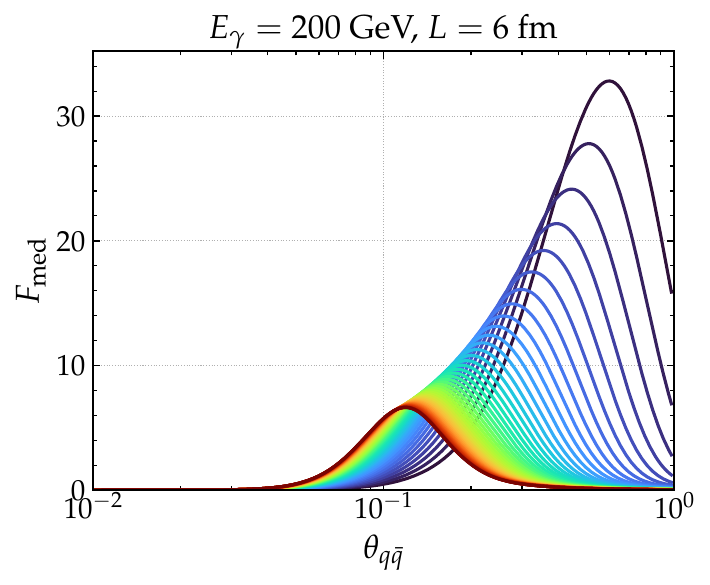}
      \includegraphics[width=0.32\textwidth]{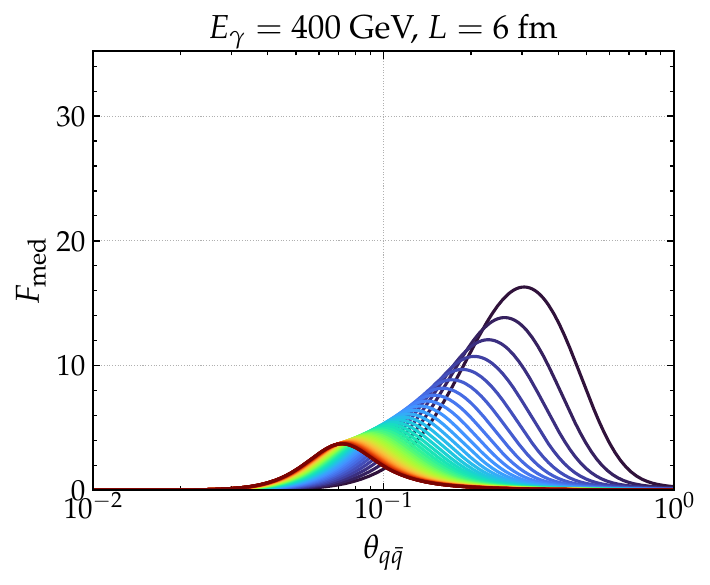}
      \includegraphics[width=0.32\textwidth]{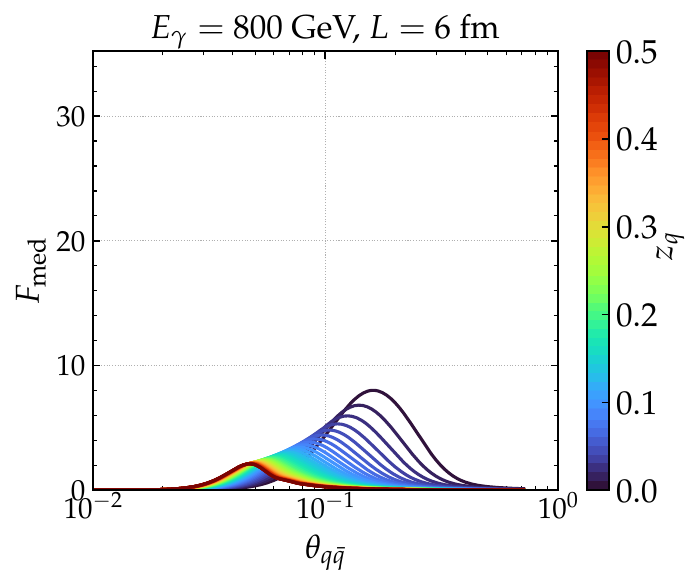}
    \caption{Numerical evaluation of $\Fmed$, Eq.~\eqref{eq:fmedDef}, as a function of the opening angle of the antenna, $\theta_{q\bar q}$, and the energy sharing fraction, $z_q$, for different photon energies and medium lengths. Note the different limits on the $y$-axis for each row.}
    \label{fig:Fmeds}
\end{figure}

In this appendix we collect some numerical results that complement those presented in the main text.

Figure~\ref{fig:Fmeds} shows the behaviour of $\Fmed$ in the parameter space described in the main text. This function controls both the total rate of emissions (see Eq.~\eqref{eq:cool-result}) and the denominator of $\TildeDeltamed$ (see Eq.\eqref{eq:tilde-delta-med}). In contrast to $\TildeDeltamed$, we note that $\Fmed$ is not bounded by one and it grows when decreasing $E_\gamma$ or, equivalently, when increasing the antenna formation time. This implies a stronger medium modification of the spectrum for long antenna formation times, as expected. 
Something that has not been discussed previously in the literature are the oscillations that we observe for $L=2$ fm and for $L=4$ fm when setting $E_\gamma=800$ GeV. 
One could argue that the former is outside the regime of validity of our approximations, since the medium is too dilute. The latter, however, is consistent with them. 
As discussed in the main text, the origin of these oscillations is an interplay between the frequency of the trigonometric functions when increasing $E_\gamma$, and the slower exponential decay of the Wilson-line correlators when decreasing $\hat q L$. 
Furthermore, we have verified that these oscillations are not an artifact of the numerical integration (see footnote \ref{fn:oscillations}).
Regarding the asymptotic behavior of $\Fmed$, we find that collinear splittings remain vacuum-like ($\Fmed=0$) for all values of $z_q$. In the large-angle regime, emissions with $z_q\sim 0.5$ are also vacuum-like. 
The small-$z_q$ curves do not vanish at wide-angles but, as we have already mentioned before, this part of the phase-space is beyond the regime of validity of our approximations and so these observations should be taken with care. 

\begin{figure}
    \includegraphics[width=0.32\textwidth]{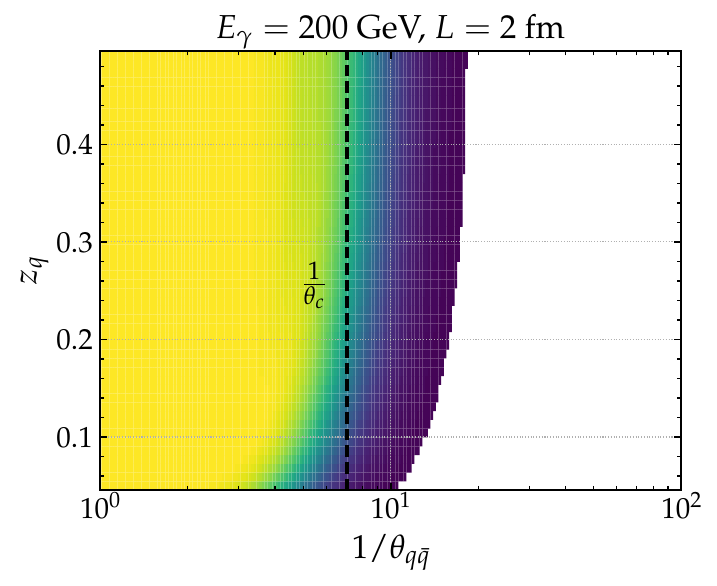}
      \includegraphics[width=0.32\textwidth]{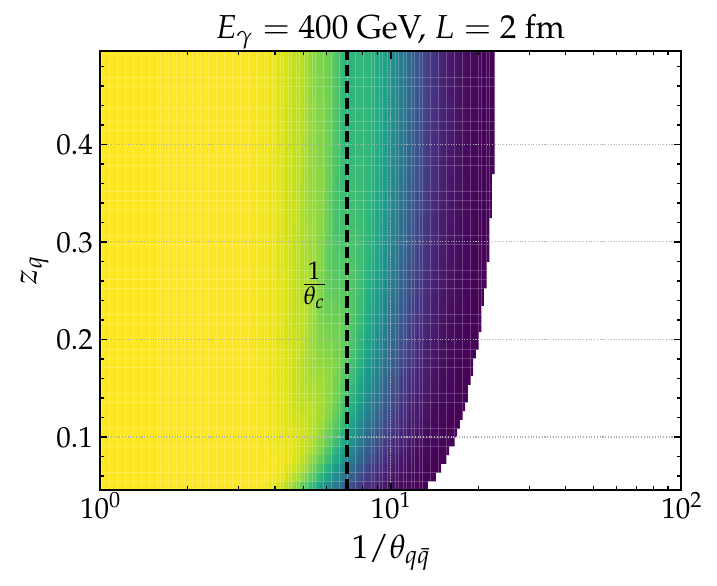}
      \includegraphics[width=0.32\textwidth]{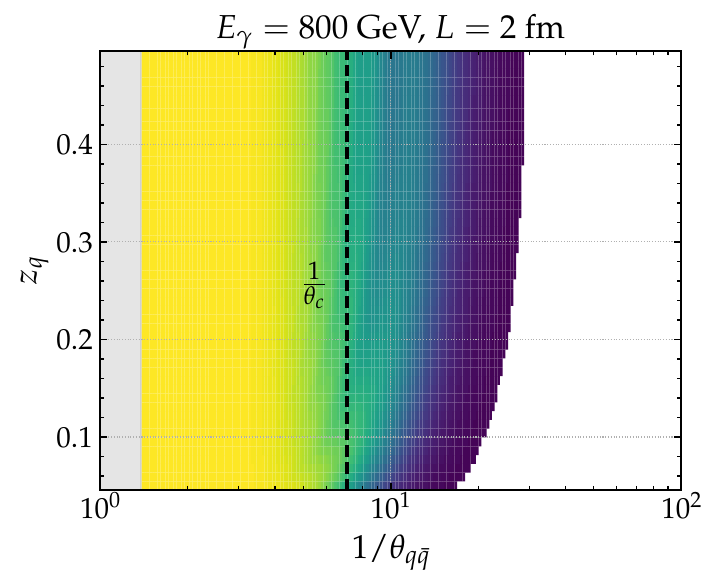}\\
          \includegraphics[width=0.32\textwidth]{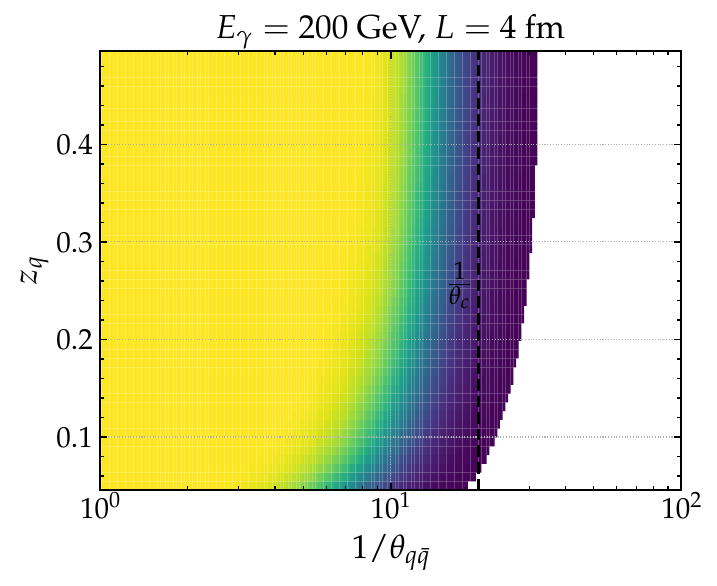}
      \includegraphics[width=0.32\textwidth]{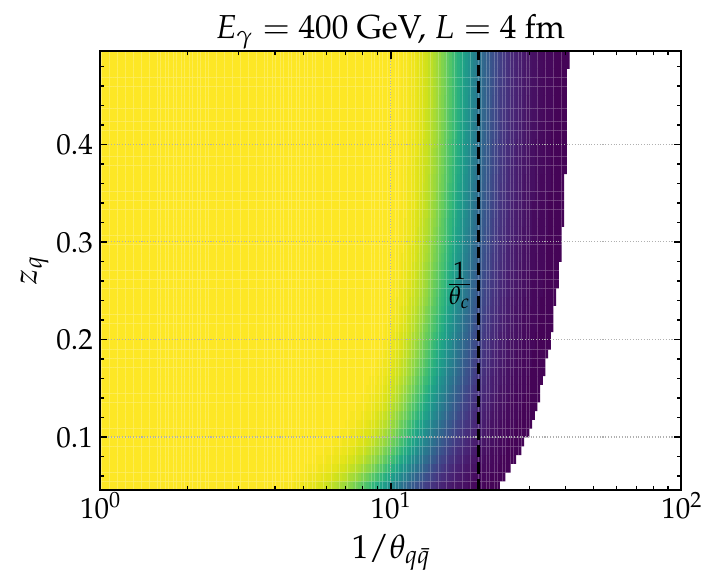}
      \includegraphics[width=0.32\textwidth]{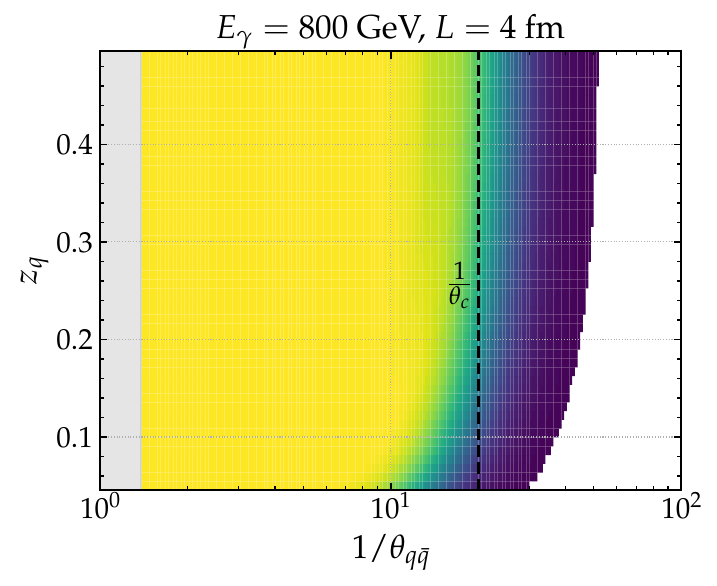}
      \\
      \includegraphics[width=0.32\textwidth]{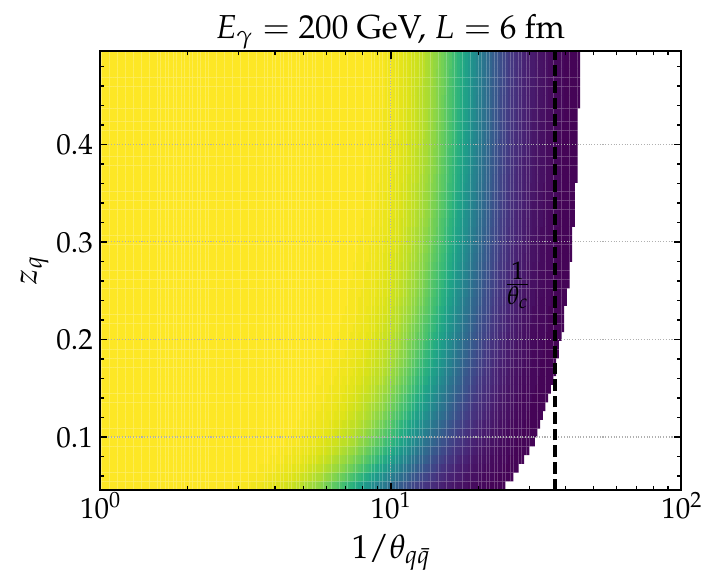}
      \includegraphics[width=0.32\textwidth]{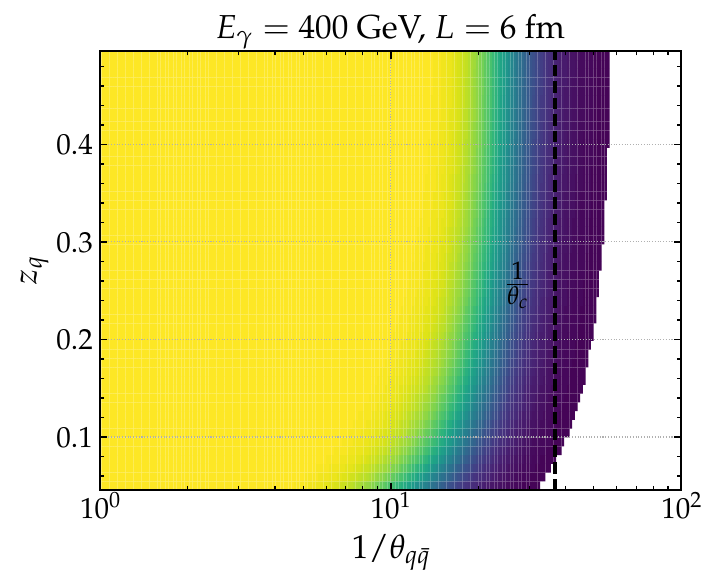}
    \includegraphics[width=0.32\textwidth]{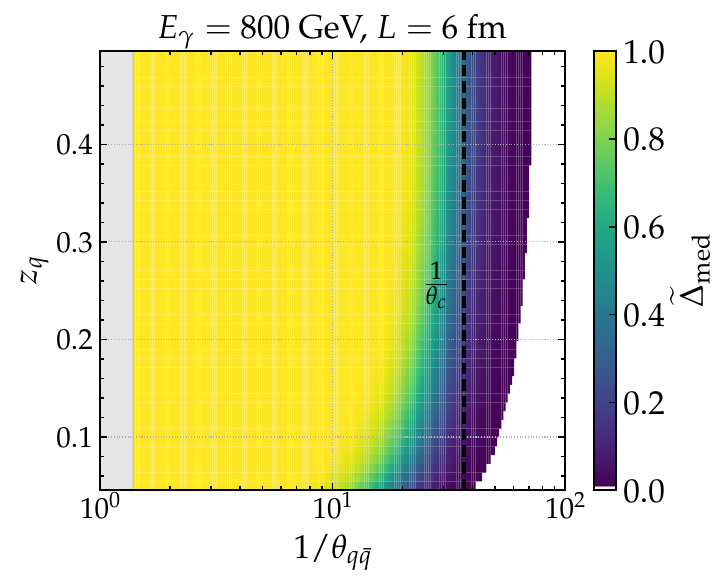}
    \caption{Lund plane representation of the generalized decoherence factor $\TildeDeltamed$ for different medium lengths and photon energies. The gray band delineates the regime where we found numerically instabilities. The vertical dashed line indicates the position of $\theta_c$, i.e.,~the value at which $\Deltamed=1-e^{-1}$.}
    \label{fig:LPs}
\end{figure}

Finally, we show in Fig.~\ref{fig:LPs} a Lund plane representation of $\TildeDeltamed$ for the parameter space explored in this work. This provides a more visual way of identifying the regions of phase-space for which certain phenomena take place. For example, the regime of total colour decoherence is clearly visible when $\theta_{q\bar q}\to 1$ for any value of $z_q$. When the antenna is sufficiently collinear, $\TildeDeltamed$ vanishes and no anti-angular ordered emissions take place. For moderate angles, we observe a strong dependence of $\TildeDeltamed$ on both medium parameters and the photon energy, as discussed in the main text. In particular, focusing on the $E_\gamma=400$ column we note how the region where $\TildeDeltamed\to1$ moves towards smaller angles when increasing $\hat qL$, thus indicating a stronger modification of the antenna emission pattern when considering denser media.

\bibliographystyle{elsarticle-num}
\bibliography{references.bib}
\end{document}